# The theoretical study on intermittency and propagation of geodesic acoustic mode in L- mode discharge near tokamak edge

Zhaoyang LIU (刘朝阳)[1,*], Yangzhong ZHANG (章扬忠)[2], Swadesh Mitter MAHAJAN [3],

Adi LIU (刘阿娣)[1], Tao XIE (谢涛)[4], Chu ZHOU (周楚)[1], Tao LAN (兰涛)[1],

Jinlin XIE (谢锦林)[1], Hong LI (李弘)[1], Ge ZHUANG (庄革)[1], and Wandong LIU (刘万东)[1]

[1] School of Nuclear Science and Technology, University of Science and Technology of China, Hefei, Anhui 230026, People's Republic of China

[2] Center for Magnetic Fusion Theory, Chinese Academy of Sciences, Hefei, Anhui 230026, People's Republic of China

[3] Institute for Fusion Studies, University of Texas at Austin, Austin, Texas 78712, USA

[4] School of Science, Sichuan University of Science and Engineering, Zigong, Sichuan 643000, People's Republic of China

[*]E-mail of corresponding author: lzy0928@mail.ustc.edu.cn

## Abstract

Through a systematically developed theory, we demonstrate that the motion of instanton identified in [Y. Z. Zhang, Z. Y. Liu, T. Xie, S. M. Mahajan, and J. Liu, Physics of Plasmas **24**, 122304 (2017)] is highly correlated to the intermittent excitation and propagation of geodesic acoustic mode (GAM) that are observed in tokamaks. While many numerical simulations have observed the phenomena, it is the first theory that reveals the physical mechanism behind GAM intermittent excitation and propagation. The preceding work is based on the micro-turbulence associated with toroidal ion temperature gradient (ITG) mode, and slab-based phenomenological model of zonal flow. When full toroidal effect are introduced into the system, two branches of zonal flow emerge: the torus-modified low frequency zonal flow (TLFZF), and GAM, necessitating a unified exploration of GAM and TLFZF. Indeed, we observe that the transition (decay) from the caviton to instanton is triggered by a rapid zero-crossing of radial group velocity of drift wave and is found to be strongly correlated with the GAM onset. Many features peculiar to intermittent GAMs, observed in real machines, are thus identified in the numerical experiment. The results will be displayed in figures and in a movie; first for single central rational surface, and then with coupled multiple central rational surfaces. The periodic bursting first shown disappears as being replaced by irregular one, more similar to the intermittent characteristics observed in GAM experiments.

## 1. Introduction

In this paper we construct a possible theoretical-computational pathway for the intermittent excitation of geodesic acoustic mode (GAM), observed, routinely, on several tokamaks such as ASDEX [1], T-10 [2], JFT-2M [3,4], HL-2A [5,6], DIII-D [7], JET [8,9], EAST [10]. Experimentally, the GAM is an intermittent, random, discrete temporal structure. More specifically, the GAM in the frequency range 10-20 KHz does not last long; typically, it lasts a few milliseconds (*e.g.*, 0.5-5ms) before disappearance, and then reappears in a shorter period of time. The entire cycle - the intermittency period - is less than 1ms near the plasma edge [5,6] and longer than 1ms away from the plasma edge [1,2,10]. During its occurrence, the GAM amplitude varies with no recognizable (so far) pattern.

We begin with presenting a theoretical framework for extending the framework of [11] – the zonal flow - drift wave system [11] – to include GAM (the relevance of ITG to GAM excitation in low mode discharge near tokamak edge has been discussed in Appendix A). 'Extending' simply means that much of the content developed in [11] will be reused in this paper, *i.e.*, the knowledge of toroidal ITG mode in two scales. In meso-scale it is the ITG wave energy equation modulated by zonal flow. In micro-scale it is the toroidal ITG eigenmode. The 2D linear mode structure pertaining to a single central rational surface is shown in figure 1. The mode structure is required for calculating Reynolds stress and group velocity in zonal flow equation and ITG wave energy equation respectively.

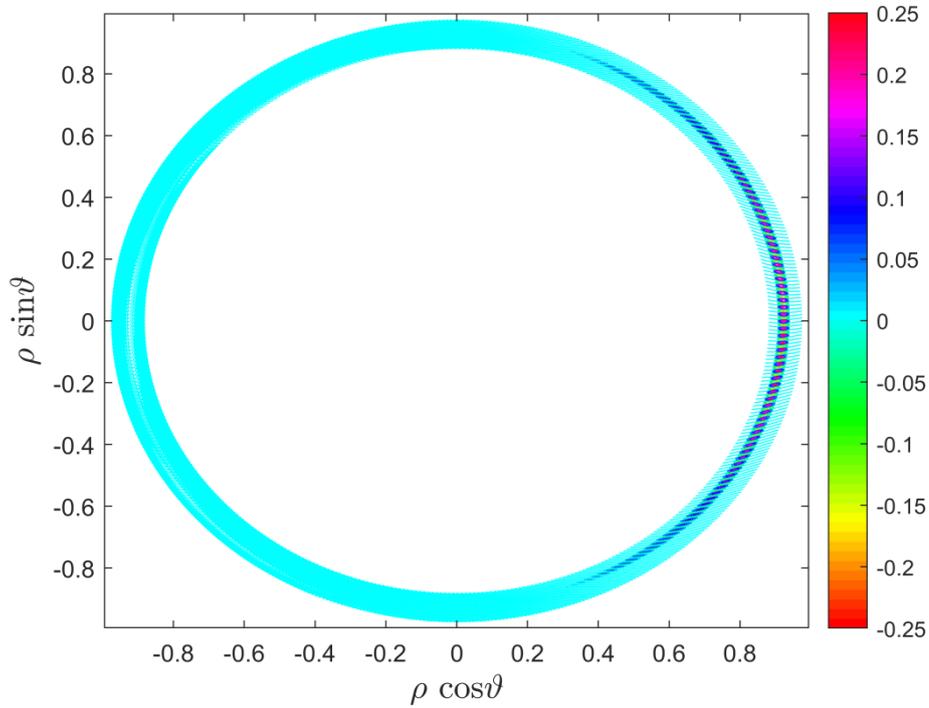

Figure 1. 2D mode structure of toroidal ITG used in this paper, calculated according to the theory of [11] and the parameters of this paper, where $\rho \equiv r/a$, $r$ is the radial position, $a$ is the minor radius, $\vartheta$ is the poloidal angle.

Two toroidal features of the ITG mode are clearly seen from figure 1, (1) The mode is ballooned towards bad curvature side; and (2) the radial scale is much broader than the poloidal one, implying quite a few sidebands are coupled to the central rational surface.

While the framework of ITG can be reused, however, the zonal flow equation in [11] should be greatly modifed. Toroidicity has two effects on the zonal flow equation. On one hand, it augments the screening factor from slab Hasegawa-Wakatani model [12-15], depending on the collisionality; on the other it causes coupling to sinusoidal sound wave through geodesic curvature. A free parameter $a_{neo}$ is phenomenologically introduced in [11] to address the first issue. In order to resolve the second issue, one ought to derive the zonal flow equation in tokamak configuration, which is precisely one task of this paper. In summary, we will reuse all microscopic scale related formalism, including linear toroidal ITG eigenmode equations, ballooning solutions for both eigenvalue and 2D mode structure, the derived Reynolds stress and group velocity in [11]. It is also important to mention that the wave energy equation Eq.(9) and its eikonal solution Eq.(13) in [11] will be reused in this work as Eq.(20); what will be different from [11] is the zonal flow equation Eq.(14), the model in meso-scale, which should be modified to incorporate toroidicity. To this purpose two basic moment equations (charge and particle conservation) for the axisymmetric mode (a synonym of zero toroidal number mode) is derived in Appendix B. Based on these two equations the toroidal zonal flow equation set is derived in section 2. It consists of two coupled equations, the zonal flow ($\bar{\upsilon}$) equation Eq.(18), (now it is coupled to the first harmonic sinusoidal component of sound wave due to geodesic curvature) and Eq.(19) for sound wave ($\chi_1^{(s)}$). These two equations can be merged into one single equation in terms of zonal flow ($\bar{\upsilon}$) as shown in Appendix D, where the explicit coupling form of the two branches TLFZF (with $a_{neo} := 1 + 2q^2$, $q$ is the safety factor of tokamak) and GAM is clearly seen. The basic form of the paper, *i.e.*, the three coupled equations Eqs.(18-20), is fully described in section 2. In addition, the physics of decay from a pair of caviton into instantons and the role played by radial group velocity crossing zero is explained briefly at the end of section 2.

The set of three equations (18-20) represents a well-posed initial value problem under specific boundary conditions as shown in section 3. The numerical methods, though essentially the same as in [11], are described in detail so that this paper is self-contained. In section 4, use is made of figures and one movie to discuss leading characteristics of GAM generated by ITG for single central rational surface. The GAM onset looks like periodic bursting. In section 5 the coupling between multiple central rational surfaces is investigated with arbitrarily chosen initial phase. The inclusion of coupling makes the periodic bursting disappear, and the resulting response looks more like the intermittent excitation observed in tokamaks. The physics regarding GAM propagation is studied systematically in section 6; its pattern is highly correlated with motion of instantons and dependent on the sign of weak dispersive media. In section 7, summary and discussion of our findings are presented. In Appendix A, we critically examine what may be the most likely source of micro-turbulence generating zonal flows-GAM phenomenon. Data on GAM experiments from nine discharges on 7 machines were collected and analyzed. Because of high collisionality, the well-known collisionless trapped electron mode (TEM or simply CTEM), and the dissipative trapped electron mode (DTEM) are excluded, from consideration. We also provide a few examples of experimental data showing that GAM could occur in ITG unstable region. Appendix B is devoted to a derivation of the charge and particle conservation equations for axisymmetric electrostatic mode; these equations constitute the basics for understanding the close relationship between TLFZF and GAM. A somewhat technical calculation for poloidal moments of Reynolds stress is given in Appendix C. In Appendix D, it is shown that, for the low frequency branch, TLFZF is consistent with Eq.(14) of [11], where the free parameter is determined to be $a_{\text{neo}} := 1 + 2q^2$; for the high frequency branch, it is consistent with the GAM dispersion in Fourier representation of [16].

## 2. The zonal flow equation set in tokamak

In the existing literature, the LFZF equation and equation for GAM are derived separately. However, we will soon show (based on the Braginskii two-fluid equations [17]) that these equations are simply two branches of a unified zonal flow system. We will assume the geometry of concentric circular magnetic surfaces in a toroidal coordinate system. Such a simplified framework should be sufficient for the exploration of the qualitative features of GAM intermittency. The starting point of the investigation is the set of two coupled conservation equations for the axisymmetric electrostatic mode [18], derived in Appendix B.

In the toroidal coordinate system $(r, \vartheta, \varsigma)$, where $r, \vartheta, \varsigma$ are respectively the radial, poloidal and toroidal coordinate, the charge conservation and particle conservation equations are (see Eqs. (B.17-18)):

$$\left( \frac{\partial}{\partial \vartheta} - \frac{r}{R} \sin \vartheta \right) \frac{\sigma_\delta}{q^2} \frac{\partial}{\partial \vartheta} \left( \bar{\underset{\sim}{n}} - \bar{\underset{\sim}{\varphi}} \right) - 2\rho_s (1 + \tau_i) \sin \vartheta \frac{\partial \bar{\underset{\sim}{n}}}{\partial r} \\ - \frac{R \rho_s^2}{c_s} \frac{\partial}{\partial t} \frac{\partial^2}{\partial r^2} \bar{\underset{\sim}{\varphi}} + \mu_B \frac{R \rho_s^2}{c_s} \frac{\partial^4}{\partial r^4} \left( \bar{\underset{\sim}{\varphi}} + \tau_i \bar{\underset{\sim}{n}} \right) = \frac{R \rho_s}{c_s^2} \nabla \cdot \bar{\underset{\sim}{u}}_{\text{NP}} \qquad (1)$$

and

$$\frac{R^2}{c_s^2} \frac{\partial^2 \bar{\underset{\sim}{n}}}{\partial t^2} - \frac{(1 + \tau_i)}{q^2} \left( \frac{\partial}{\partial \vartheta} - \frac{r}{R} \sin \vartheta \right) \frac{\partial \bar{\underset{\sim}{n}}}{\partial \vartheta} - \frac{2 R \rho_s}{c_s} \sin \vartheta \frac{\partial}{\partial t} \frac{\partial}{\partial r} \left( \bar{\underset{\sim}{\varphi}} + \tau_i \bar{\underset{\sim}{n}} \right) \\ - \frac{R^2 \rho_s^2}{c_s^2} \frac{\partial^2}{\partial t^2} \frac{\partial^2}{\partial r^2} \bar{\underset{\sim}{\varphi}} + \mu_B \frac{R^2 \rho_s^2}{c_s^2} \frac{\partial}{\partial t} \frac{\partial^4}{\partial r^4} \left( \bar{\underset{\sim}{\varphi}} + \tau_i \bar{\underset{\sim}{n}} \right) = \frac{R^2 \rho_s}{c_s^3} \frac{\partial}{\partial t} \nabla \cdot \bar{\underset{\sim}{u}}_{\text{NP}} \qquad (2)$$

In Eq.(1) $\sigma_\delta \equiv 2 m_i c_s / m_e \nu_{ei} R$ is the normalized parallel conductivity, $R$ is the major radius, $c_s^2 \equiv T_e / m_i$, $T_e$ ($T_i$) is the equilibrium electron (ion) temperature, $m_e$ ($m_i$) is the electron (ion) mass, $\rho_s \equiv c_s / \omega_{ci}$ ($\rho_i \equiv \sqrt{T_i / m_i} / \omega_{ci}$) is the ion Larmor radius at electron (ion) temperature, $v_{ei}$ ($v_{ii}$) is the electron-ion (ion-ion) collision frequency, $\mu_B \equiv 3v_{ii}\rho_i^2/10$ is the classical perpendicular viscosity coefficient, $\tau_i \equiv T_i/T_e$. $\bar{\underset{\sim}{n}}$ ($\bar{\underset{\sim}{\varphi}}$) is the normalized density (electrostatic potential) fluctuation associated with the axisymmetric mode in meso-scale, $\bar{\underset{\sim}{\boldsymbol{u}}}_{NP} \equiv \left\langle \rho_s^2 c_s^2 \left(\boldsymbol{b}\times\nabla\tilde{\phi}\cdot\nabla\right)\nabla_\perp \tilde{\phi} \right\rangle_{en}$ is the ensemble average of nonlinear polarization drift, $\tilde{\phi}$ is the normalized electrostatic potential of micro-turbulence such as drift waves, $\langle...\rangle_{en}$ stands for ensemble average over microscopic scale.

Due to the large parallel conductivity (e.g., $\sigma_\delta > 10^2$) in tokamak plasmas, the leading term in Eq. (1) is

$$\left(\frac{\partial}{\partial\vartheta} - \frac{r\sin\vartheta}{R}\right)\frac{\sigma_\delta}{q^2}\frac{\partial}{\partial\vartheta}\left(\bar{\underset{\sim}{n}} - \bar{\underset{\sim}{\varphi}}\right) = 0 \tag{3}$$

For this study, it is natural to split $\bar{\underset{\sim}{\varphi}}$ and $\bar{\underset{\sim}{n}}$ into two parts: the zonal density-flow part ($\bar{n} - \bar{\varphi}$, no poloidal dependence) and the part ($\underset{\sim}{n} - \underset{\sim}{\varphi}$) that depends on the poloidal angle [18]:

$$\bar{\underset{\sim}{n}}(r,\vartheta,t) = \bar{n}(r,t) + \underset{\sim}{n}(r,\vartheta,t), \quad \bar{\underset{\sim}{\varphi}}(r,\vartheta,t) = \bar{\varphi}(r,t) + \underset{\sim}{\varphi}(r,\vartheta,t). \tag{4}$$

Eq.(3), containing only $\vartheta$ derivatives, will require $\underset{\sim}{n} = \underset{\sim}{\varphi}$; the leading order response is, thus, *adiabatic*.

Averaging over the magnetic surface [18] ($\varepsilon \equiv r/R \ll 1$, the inverse aspect ratio)

$$\langle...\rangle_\vartheta \equiv \oint d\vartheta ...(1+\varepsilon\cos\vartheta) \tag{5}$$

on Eq.(1) yields

$$2\rho_s(1+\tau_i)\frac{\partial}{\partial r}\langle\sin\vartheta\underset{\sim}{\varphi}\rangle_\vartheta + \frac{R\rho_s^2}{c_s}\frac{\partial}{\partial t}\frac{\partial^2}{\partial r^2}\left(\bar{\varphi}+\langle\underset{\sim}{\varphi}\rangle_\vartheta\right)$$
$$+R\rho_s^3\langle\Pi(r,\vartheta)\rangle_\vartheta - \mu_B\frac{R\rho_s^2}{c_s}\frac{\partial^4}{\partial r^4}\left[\bar{\varphi}+\tau_i\bar{n}+(1+\tau_i)\langle\underset{\sim}{\varphi}\rangle_\vartheta\right] = 0 \tag{6}$$

In Eq.(6) $\Pi(r,\vartheta) \equiv \nabla\cdot\left[\left(\boldsymbol{b}\times\nabla\tilde{\phi}\cdot\nabla\right)\nabla_\perp\tilde{\phi}\right]$. The symbol $\langle...\rangle_{en}$ can be removed, because the surface averaging is equivalent to emsemble average, which only removes micro-scale quantities.

As long as $\Pi(r,\vartheta)$ appears in the integral over $\vartheta$, the definition is justified. Noticeably, this step annihilates the leading term of Eq.(1); the leftover Eq.(6) is the charge conservation equation for an axisymmetric mode that contains a geodesic curvature induced coupling to the sinusoidal wave. It will be further simplified after discussing some relations arising from particle conservation, Eq.(2).

The magnetic surface average of Eq.(2) yields

$$\frac{R^2}{c_s^2}\frac{\partial^2}{\partial t^2}\left(\bar{n}+\langle\bar{\varphi}\rangle_\vartheta\right) - \frac{2R\rho_s}{c_s}\frac{\partial}{\partial t}\frac{\partial}{\partial r}(1+\tau_i)\langle\sin\vartheta\tilde{\varphi}\rangle_\vartheta - \frac{R^2\rho_s^2}{c_s^2}\frac{\partial^2}{\partial t^2}\frac{\partial^2}{\partial r^2}\left(\bar{\varphi}+\langle\bar{\varphi}\rangle_\vartheta\right)$$
$$+\mu_B\frac{R^2\rho_s^2}{c_s^2}\frac{\partial}{\partial t}\frac{\partial^4}{\partial r^4}\left[\bar{\varphi}+\tau_i\bar{n}+(1+\tau_i)\langle\tilde{\varphi}\rangle_\vartheta\right] = \frac{R^2\rho_s^3}{c_s}\frac{\partial}{\partial t}\langle\Pi(r,\vartheta)\rangle_\vartheta \quad (7)$$

By acting $(R/c_s)(\partial/\partial t)$ on Eq.(6) and adding the result to Eq.(7), a straightforward algebra leads to a rather simple equation $(\partial^2/\partial t^2)\left(\bar{n}+\langle\tilde{\varphi}\rangle_\vartheta\right)=0$. We choose the trivial solution, unless some greater constant zonal density is observed

$$\bar{n} = -\langle\tilde{\varphi}\rangle_\vartheta = -\varepsilon\oint d\vartheta\cos\vartheta\tilde{\varphi}. \quad (8)$$

Eq.(8) implies that the zonal density arises from the first cosinoidal component of $\tilde{\varphi}$, however, at the order of $O(\varepsilon)$.

Eq.(8) is now used to eliminate $\bar{n}$ from Eq.(2) to yield

$$\frac{R^2}{c_s^2}\frac{\partial^2}{\partial t^2}\left(\tilde{\varphi}-\varepsilon\oint d\vartheta\cos\vartheta\tilde{\varphi}\right) - \frac{(1+\tau_i)}{q^2}\left(\frac{\partial}{\partial\vartheta}-\varepsilon\sin\vartheta\right)\frac{\partial\tilde{\varphi}}{\partial\vartheta}$$
$$-\frac{2R\rho_s}{c_s}\sin\vartheta\frac{\partial}{\partial t}\frac{\partial}{\partial r}\left(\bar{\varphi}+(1+\tau_i)\tilde{\varphi}-\tau_i\varepsilon\oint d\vartheta\cos\vartheta\tilde{\varphi}\right)$$
$$+\mu_B\frac{R^2\rho_s^2}{c_s^2}\frac{\partial}{\partial t}\frac{\partial^4}{\partial r^4}\left(\bar{\varphi}+(1+\tau_i)\tilde{\varphi}-\tau_i\varepsilon\oint d\vartheta\cos\vartheta\tilde{\varphi}\right) \quad (9)$$
$$= \frac{R^2\rho_s^2}{c_s^2}\frac{\partial}{\partial t}\left[\frac{\partial}{\partial t}\frac{\partial^2}{\partial r^2}\left(\bar{\varphi}+\tilde{\varphi}\right)+\rho_s c_s\Pi(r,\vartheta)\right]$$

For meso-scale ($\rho_s(\partial/\partial r)\ll 1$) solution all terms containing $\rho_s$ can be neglected in Eq. (9). It becomes

$$\frac{R^2}{c_s^2}\frac{\partial^2}{\partial t^2}\left(\underset{\sim}{\varphi} - \varepsilon\oint d\vartheta \cos\vartheta\underset{\sim}{\varphi}\right) - \frac{(1+\tau_i)}{q^2}\left(\frac{\partial}{\partial\vartheta} - \varepsilon\sin\vartheta\right)\frac{\partial\underset{\sim}{\varphi}}{\partial\vartheta} = 0. \tag{10}$$

Since the coefficients of Eq.(10) depends only on $\vartheta$, $\underset{\sim}{\varphi}$ can be separated as

$$\underset{\sim}{\varphi}(r,\vartheta,t) := \chi_\nu^{(\alpha)}(r,t) F_\nu^{(\alpha)}(\vartheta). \tag{11}$$

Interestingly, if we further split the $\vartheta$ dependence as $F_\nu^{(\alpha)}(\vartheta) := \exp(-\varepsilon\cos\vartheta/2) f_\nu^{(\alpha)}(\vartheta)$, the homogeneous part of Eq.(10) can be cast into the canonical Mathieu equation [19]

$$\frac{d^2 f_\nu^{(\alpha)}}{dz^2} + \left(4\lambda_\nu^{(\alpha)} - 2\varepsilon\cos 2z\right) f_\nu^{(\alpha)} = 0 \tag{12}$$

In Eq.(12) $\vartheta = 2z + \pi$, $\lambda_\nu^{(\alpha)} \equiv q^2 R^2 \omega^2 / (1+\tau_i) c_s^2$, $\alpha$ and $\nu$ are characteristic exponents, $\omega$ is the Fourier transform of time $\partial/\partial t \to -i\omega$. The inhomogeneous part of Eq.(10) only contributes to the cosinoidal component of $\underset{\sim}{\varphi}$. Characteristic functions and characteristic values of Eq.(10) are listed in Table 1.

Table 1. Characteristic functions and values of Eq.(10)

|  | Characteristic functions | Characteristic values |
| --- | --- | --- |
| 1st sine | $f_1^{(s)} = \sin\vartheta + \dfrac{\varepsilon}{12}\sin 2\vartheta$ | $\lambda_1^{(s)} = 1 + O(\varepsilon^2)$ |
| 1st cosine | $F_1^{(c)} = \cos\vartheta - \dfrac{\varepsilon}{6}\cos 2\vartheta$ | $\lambda_1^{(c)} = 1 + O(\varepsilon^2)$ |
| 2nd sine | $f_2^{(s)} = \sin 2\vartheta + \varepsilon\left(\dfrac{1}{20}\sin 3\vartheta - \dfrac{1}{12}\sin\vartheta\right)$ | $\lambda_2^{(s)} = 4 + O(\varepsilon^2)$ |

Substituting these characteristic functions into Eq.(9) yields

$$\frac{R^2}{c_s^2}\frac{\partial^2 \chi_\nu^{(\alpha)}}{\partial t^2}\left(F_\nu^{(\alpha)} - \varepsilon\oint d\vartheta \cos\vartheta F_\nu^{(\alpha)}\right) + \frac{(1+\tau_i)}{q^2}\lambda_\nu^{(\alpha)}\chi_\nu^{(\alpha)}\left(F_\nu^{(\alpha)} - \varepsilon\oint d\vartheta \cos\vartheta F_\nu^{(\alpha)}\right)$$
$$-\frac{2R\rho_s}{c_s}\sin\vartheta\frac{\partial}{\partial t}\frac{\partial}{\partial r}\left\{\overline{\varphi} + \left[(1+\tau_i)F_\nu^{(\alpha)} - \tau_i\varepsilon\oint d\vartheta \cos\vartheta F_\nu^{(\alpha)}\right]\chi_\nu^{(\alpha)}\right\}$$
$$+\mu_B\frac{R^2\rho_s^2}{c_s^2}\frac{\partial}{\partial t}\frac{\partial^4}{\partial r^4}\left\{\overline{\varphi} + \left[(1+\tau_i)F_\nu^{(\alpha)} - \tau_i\varepsilon\oint d\vartheta \cos\vartheta F_\nu^{(\alpha)}\right]\chi_\nu^{(\alpha)}\right\}$$
$$=\frac{R^2\rho_s^2}{c_s^2}\left[\frac{\partial^2}{\partial t^2}\frac{\partial^2 \overline{\varphi}}{\partial r^2} + \frac{\partial^2}{\partial t^2}\frac{\partial^2 \chi_\nu^{(\alpha)}}{\partial r^2}F_\nu^{(\alpha)} + \rho_s c_s\frac{\partial}{\partial t}\Pi(r,\vartheta)\right]$$
.(13)

Because the Mathieu characteristic functions constitute an orthogonal complete set, multiplying Eq.(13) by $\breve{F}_\nu^{(\alpha)} \equiv \exp(\varepsilon\cos\vartheta/2)f_\nu^{(\alpha)}$ and integrating over $\vartheta$ yields the radial equation of each harmonic. For the first sinusoidal component, $\nu=1$ and $\alpha:=s$, the radial equation becomes

$$\left[\frac{R^2}{c_s^2}\frac{\partial^2}{\partial t^2} + \frac{(1+\tau_i)}{q^2}\right]\chi_1^{(s)} - \frac{R^2\rho_s^2}{c_s^2}\frac{\partial^2}{\partial t^2}\frac{\partial^2 \chi_1^{(s)}}{\partial r^2}$$
$$-\frac{2R\rho_s}{c_s}\frac{\partial}{\partial t}\frac{\partial \overline{\varphi}}{\partial r} + \mu_B(1+\tau_i)\frac{R^2\rho_s^2}{c_s^2}\frac{\partial}{\partial t}\frac{\partial^4 \chi_1^{(s)}}{\partial r^4}$$
$$=\frac{2R^2\rho_s^3}{c_s}\frac{\partial}{\partial t}\oint d\vartheta\exp(\varepsilon\cos\vartheta/2)\left(\sin\vartheta + \frac{\varepsilon}{12}\sin 2\vartheta\right)\Pi(r,\vartheta)$$ . (14)

The same substitution - $\underset{\sim}{\varphi} \to \chi_1^{(s)}(r,t)F_1^{(s)}(\vartheta)$ - can be similarly applied to the charge conservation Eq.(6), which on integration over $\vartheta$, reduces to (at lowest order)

$$\rho_s(1+\tau_i)\frac{\partial \chi_1^{(s)}}{\partial r} + \frac{R\rho_s^2}{c_s}\frac{\partial}{\partial t}\frac{\partial^2 \overline{\varphi}}{\partial r^2} - \mu_B\frac{R\rho_s^2}{c_s}\frac{\partial^4 \overline{\varphi}}{\partial r^4} + R\rho_s^3\langle\Pi(r,\vartheta)\rangle_\vartheta = 0. \quad (15)$$

In Eq.(15) the last term is

$$\oint d\vartheta\Pi(r,\vartheta) = \frac{\partial^2}{\partial r^2}\left(-\oint d\vartheta\frac{\partial\tilde{\phi}}{r\partial\vartheta}\frac{\partial\tilde{\phi}}{\partial r}\right) \quad (16)$$

is the average micro-turbulence drive that generates the zonal flow and GAM; the quantity behind the second order radial derivative in Eq.(15) is the well-known Reynolds stress.

The poloidal moment of the Reynolds stress (introduced by the toroidal coupling) is obtained upon performing surface average on Eq.(14)

$$\tilde{\mathfrak{R}}[K] \equiv -\rho_s^2 c_s^2 \oint d\vartheta \left( \frac{\partial \tilde{\phi}}{r \partial \vartheta} \frac{\partial \tilde{\phi}}{\partial r} \right) K(\vartheta). \tag{17}$$

In Eq.(17) $K(\vartheta) = 1$, $\sin\vartheta$, $\cos\vartheta$, $\sin 2\vartheta$. The analytical expression for $\tilde{\mathfrak{R}}[K]$ can be obtained straightforwardly as shown in Appendix C.

Notice that $\tilde{\mathfrak{R}}[K]$ is the functional of ITG mode amplitude. It is not a constant in time, because ITG mode amplitude is modulated by zonal flow. As long as the group velocities of the drift wave and zonal flows are real in spatiotemporal representation, the modulation of drift wave energy by zonal flow is the phase modulation via $\cos^2\Theta$ (equivalently $\cos(2\Theta)/2$, $\Theta$ is the phase (eikonal) of ITG mode amplitude) for a single rational surface as shown in [11].

Eqs.(14), (15) constitute the zonal flow-sound wave system purely based on Braginskii fluid model. The equivalent zonal flow equation pertaining to the single rational surface, Eq.(15) becomes

$$\frac{\partial \bar{\upsilon}}{\partial t} - \mu \frac{\partial^2 \bar{\upsilon}}{\partial r^2} = -\frac{\psi c_s^2}{R}(1+\tau_i)\chi_1^{(s)} - \frac{1}{2}\frac{\partial}{\partial r}\left\{ \left(\tilde{\mathfrak{R}}[1] + \varepsilon\tilde{\mathfrak{R}}[\cos\vartheta]\right)\cos 2\Theta \right\}. \tag{18}$$

In Eq.(18) $\bar{\upsilon} \equiv \rho_s c_s \partial \bar{\varphi}/\partial r$ is the zonal flow, $\psi \equiv 1 - (r-r_j)/L_{T_e}$ describes the (linear) electron temperature profile, $L_{T_e}$ is the electron temperature gradient length, $r_j$ is the radial position of rational surface. In the slab limit (by dropping $\chi_1^{(s)}$ and $\tilde{\mathfrak{R}}[\cos\vartheta]$) Eq.(18) reduces to Eq.(19) of [11] with $a_{\text{neo}} := 1$. Substituting Eq.(18) into Eq.(14) yields the equation for the first harmonic sinusoidal component of sound wave

$$\left[\frac{\partial^2}{\partial t^2}\left(1+D(\tau_e)\rho_i^2\frac{\partial^2}{\partial r^2}\right)+2(1+\tau_i)\frac{\psi c_s^2}{R^2}\left(1+\frac{1}{2q^2}\right)\right]\chi_1^{(s)}+\mu(1+\tau_i)\rho_s^2\frac{\partial}{\partial t}\frac{\partial^4\chi_1^{(s)}}{\partial r^4}$$

$$=\frac{2\mu}{R}\frac{\partial^2\bar{\upsilon}}{\partial r^2}-\frac{1}{R}\frac{\partial}{\partial r}\left\{\left(\tilde{\mathfrak{R}}[1]+\varepsilon\tilde{\mathfrak{R}}[\cos\vartheta]\right)\cos 2\Theta\right\}$$ . (19)

$$+\frac{\rho_s}{c_s}\frac{\partial}{\partial t}\frac{\partial^2}{\partial r^2}\left\{\left(\tilde{\mathfrak{R}}[\sin\vartheta]+\frac{\varepsilon}{3}\tilde{\mathfrak{R}}[\sin 2\vartheta]\right)\cos 2\Theta\right\}$$

As mentioned in the introduction that Eq.(13) in [11] will be reused, it is simply quoted here

$$\Theta(r,t)=k_\vartheta\int_0^t dt'\bar{\upsilon}\left(r-\int_{t'}^t ds\upsilon_{gr}(\vartheta(s)),t'\right)$$ [20], $d\vartheta(s)/ds=\upsilon_{gy}(\vartheta)/r_j$, (20)

with $\vartheta(s=0)=0$, $\upsilon_{gr}(\upsilon_{gy})$ is the radial (poloidal) group velocity. $\mu$ is the perpendicular viscosity; for classical fluid model, $\mu\to\mu_B\equiv 3\nu_{ii}\rho_i^2/10$. For a real plasma, the viscosity is likely to be anomalous, $\mu=a_\mu\mu_B$, $a_\mu$ being the measure of anomaly. In Eq.(20) $k_\vartheta\equiv m/r_j$ is the poloidal wave vector, and $n$ is for toroidal mode number of ITG mode. At the mode rational surface, the poloidal mode number $m=nq(r_j)$, $q(r_j)$ being the safety factor.

A crucial comment, regarding the weak dispersive - finite Larmor radius (FLR) - term $D(\tau_e)\rho_i^2(\partial^2/\partial r^2)$ in Eq.(19), is in order. In our results from Braginskii model, the dispersive coefficient $D(\tau_e)$ is proportional to $-\tau_e\equiv -T_e/T_i$. However, GAM propagation observed in experiments [21,22] seems to suggest $D(\tau_e)>0$ [23]. Various results of positive $D(\tau_e)$ have been obtained by making use of gyrokinetic theories, however, most of them does not have valid cold ion limit [16]. According to the derivation of [16] by Smolyakov *et al.*, $D(\tau_e)$ is a function of $\tau_e$. For warm ions it is positive and becomes negative in cold ion limit, the same as what we obtained. The abovementioned discrepancies in existing literature can be seen, *e.g.*, in figure 3 of [24] for the comparison of [16] with [25,26] (for gyrokinetic derivation). There are two groups working on basis of fluid model in literature, with FLR correction [27] and without FLR correction [28,29]. The FLR correction makes negative dispersive coefficient increase with growing ion temperature to positive as shown in figure 4 of [27]. Qualitatively it is very similar to $D(\tau_e)$ in [16]. Without FLR correction $D(\tau_e)$ remains negative. In this paper we just simply take the result of [16] and/or [27] as if the FLR term in the fluid model moves from cold ion to warm ion. The related GAM propagation issue will be discussed in section 6.

Eqs.(18), (19) are two components of the zonal flow equation set in tokamak. This is in contrast to [11] where only one component exists. The second component comes into the system because of the toroidal coupling to sinusoidal component of sound wave owing to the geodesic curvature. The two component set contains two branches in different frequency regime GAM and TLFZF. Interested readers may wonder if the low frequency branch of this set is consistent with the discussion in [11]. The answer to this question is 'yes' as briefly discussed in Appendix D.

Before carrying out numerical calculations, it may be appropriate to briefly describe the physics involving decay of a pair of caviton into instantons, and the role played by radial group velocity *etc.* (for details, please see Section V of [11]).

As shown in Eq. (20), the zonal flow modulates drift wave energy in phase ($\bar{\phi} \sim \cos\Theta$, where $\bar{\phi}$ is the amplitude of the drift wave). The radial group velocity $\upsilon_{gr}$ appears as the argument of $\bar{\upsilon}$ under integral, describing the movement of drift wave energy along the radial characteristic line $r_g(t) := \int_{t'}^{t} ds \upsilon_{gr}(\vartheta(s))$. According to the calculation of drift wave group velocity (ITG fluid model in [11] and in this paper), $\upsilon_{gr}$ always consists of two consecutive distinct phases: a long slowly varying part (at high level) and a short sudden spike crossing zero. Notice that the zonal flow $\bar{\upsilon}$ (in this paragraph refers to LFZF only) is localized around the region where the Reynolds stress is not small (reaction region). Before crossing zero, $\upsilon_{gr}$ is large, $r_g$ would run out of the reaction region depending on the reference position $r$, where $\bar{\upsilon}$ is too small to contribute to the integral in Eq.(20). This process corresponds to formation of a pair of caviton as $\Theta$ is slowly varying. Upon $\upsilon_{gr}$ zero-crossing, the sign is changed, making $r_g$ smaller, and pulling the local integrand $\bar{\upsilon}$ back to the reaction region, making $\bar{\upsilon}$ contribute to $\Theta$ again. Such a process occurs on different instants at different reference positions, just like wave propagation. It annihilates a pair of caviton into instantons and propagates along with $r_g$, taking $\Theta$ far away from reaction region. The details can be seen from figures 6-10 in [11], in particular figure 7.

### 3. Numerical methods for the dimensionless zonal flow equation set

We first list the various normalization introduced in [11]. The zonal flow speed is normalized to $\bar{\upsilon}_z \equiv \rho_s c_s k_\vartheta \hat{s} \sqrt{I_m(x_0)}$ ($\bar{\upsilon} = V \bar{\upsilon}_z$), time is normalized to $\omega_Z \equiv \bar{\upsilon}_z k_\vartheta$ ($t\omega_Z = \tau$), the dimensionless micro-radius is defined as $x = k_\vartheta \hat{s}(r - r_j)$, $\hat{s}$ is the magnetic shear, $I_m(x_0)$ stands for the turbulence level at the dimensionless peak position of static Reynold stress $x_0$. The zonal flow Reynolds number is defined as $R_z \equiv \omega_Z / \mu k_\vartheta^2 \hat{s}^2$ ($\mu = 3a_\mu v_{ii} \rho_i^2 / 10$), and the dimensionless Reynolds stress is $R[K] \equiv \tilde{\Re}[K] / \rho_s^2 c_s^2 k_\vartheta^2 \hat{s} I_m(x_0)$. Then we define the dimensionless first harmonic sinusoidal component of sound wave to be $\chi \equiv k_\vartheta R \chi_1^{(s)}$. By introducing two frequencies

$$\omega_G^2 \equiv 2(1+\tau_i)\frac{c_s^2}{R^2}\left(1+\frac{1}{2q^2}\right), \quad \varpi_G^2 \equiv \frac{1+\tau_i}{q^2}\frac{c_s^2}{R^2}, \tag{21}$$

and $\hat{\omega}_G \equiv \omega_G / \omega_Z$, $\hat{\varpi}_G \equiv \varpi_G / \omega_Z$, Eqs.(18-20) can be cast into the dimensionless form

$$\frac{\partial V}{\partial \tau} - \frac{1}{R_z}\frac{\partial^2 V}{\partial x^2} = -\psi q^2 \hat{\varpi}_G^2 \chi - \frac{1}{2}\frac{\partial}{\partial x}\left\{\left(R[1] + \varepsilon R[\cos\vartheta]\right)\cos 2\Theta\right\}, \tag{22}$$

$$\Theta(x,\tau) = \int_0^\tau d\tau' V\left(x - \hat{s}\int_{\tau'}^\tau d\xi \hat{\upsilon}_{gr}(\vartheta(\xi)), \tau'\right), \quad \frac{d\vartheta(\xi)}{d\xi} = \hat{\upsilon}_{gy}(\vartheta(\xi))/(k_\vartheta r_j), \tag{23}$$

- 14 -

$$\left[\frac{\partial^2}{\partial \tau^2}\left(1+D(\tau_e)\delta^2\frac{\partial^2}{\partial x^2}\right)+\psi\hat{\omega}_G^2+(1+\tau_e)\frac{\delta^2}{R_z}\frac{\partial}{\partial \tau}\frac{\partial^4}{\partial x^4}\right]\chi = \frac{2}{R_z}\frac{\partial^2 V}{\partial x^2} -$$
$$\frac{\partial}{\partial x}\{(\text{R}[1]+\varepsilon\text{R}[\cos\vartheta])\cos 2\Theta\} + c_\text{R}\frac{\partial}{\partial \tau}\frac{\partial^2}{\partial x^2}\left\{\left(\text{R}[\sin\vartheta]+\frac{\varepsilon}{3}\text{R}[\sin 2\vartheta]\right)\cos 2\Theta\right\} \quad , (24)$$

where $\hat{\upsilon}_{gr} \equiv \upsilon_{gr}/\bar{\upsilon}_z$, $\hat{\upsilon}_{gy} \equiv \upsilon_{gy}/\bar{\upsilon}_z$, $\delta \equiv \rho_i k_\vartheta \hat{s}$, $c_\text{R} \equiv R k_\vartheta \hat{s} \omega_Z / \omega_{ci}$. Before solving the preceding equations, the Reynolds stress (in Appendix C) and the ITG wave group velocities (Eqs.(10-11) in [11]) are calculated by making use of the 2D mode structure (displayed in figure 1) of weakly asymmetric ballooning theory (WABT [30,31]).

The solution of the initial value problem is worked out for the initial conditions: $V(x,0) = 0$, $\Theta(x,0) = 0$ and $\chi(x,0) = 0$. Since the phase modulation of drift wave energy is significant only inside the so-called reaction region, where the Reynolds stress is not small, the boundary condition for drift wave energy is not important; the latter is relevant only in the instanton phase that has a vanishing coupling to zonal flow outside reaction region. However, the signal of phase function $\Theta$ could propagate to a place far away from the reaction region denoted by $x_{\pm\infty}$, where the cut-off has been introduced ($\Theta(x_{\pm\infty}, \tau) = 0$) [11].

The boundary condition for the zonal flow equation set has to be set up differently with respect to left and right side in contrast to [11], since the data of GAM are measured not too far away from plasma edge, and edge effects on GAM could be important. On the right side, the zero Dirichlet (reflecting) boundary condition is chosen at the plasma edge for $V(x_{\text{edge}}, \tau) = 0$, $\chi(x_{\text{edge}}, \tau) = 0$, where $x_{\text{edge}}$ denotes the position of plasma edge ($r = a$). On the left side, an absorptive boundary condition is set up [32]. It sits far away behind turning points.

The dimensionless set of Eqs.(22-24), combined with the assumed boundary conditions, constitutes a well-posed initial value problem, which is solved by making use of the finite difference methods, where the spatiotemporal grids are discretized as $(r_k, t_m)$,

$r_k \equiv r_{j,-12} + k \cdot \Delta r$, $t_m \equiv m \cdot \Delta t$, k=0,1,...,K and m = 0,1,...,M are integers. In this paper we choose K = 512, M = 12000, $\Delta r = (a - r_{j,-12})/K$, $\Delta t = 5\mu s$, where $r_{j,\ell} \equiv r_j + (x_0 + \ell\sqrt{n\sigma})/|k_\vartheta \hat{s}|$, $\ell = 0, \pm 1, ..., \pm 12$, $r_{j,0}$ stands for the peak position of static Reynold stress (the region from $r_{j,-2}$ to $r_{j,2}$ is more or less the reaction region) and $\sqrt{n\sigma}/|k_\vartheta \hat{s}|$ stands for the half width of the Gaussian peak (the analytic formulae of static Reynold stress are given in Eqs.(C.26-C.28)). The dimensionless spatiotemporal step sizes are $\Delta x \equiv |k_\vartheta \hat{s}| \cdot \Delta r$ and $\Delta \tau \equiv \omega_Z \cdot \Delta t$ respectively.

The zonal flow equation set Eqs.(22), (24) is solved by making use of the 2$^{nd}$ order Crank-Nicolson method [33] with accuracy up to $6 \times 10^{-3}$ for $\omega_G \approx 15\text{kHz}$. The wave energy equation Eq.(23) is directly integrated as shown in Appendix C of [11].

For illustrative purposes the numerical experiment is performed for the parameters corresponding to the shot #141958 on DIII-D [7] shown in Table 2 (where $\bar{\eta}_i \equiv (1+\eta_i)/\tau_e$, $\eta_i \equiv L_n/L_{T_i}$, $\varepsilon_n \equiv L_n/R$, $L_n$ ($L_{T_i}$) is the density (ion temperature) gradient length, $B$, $N_e$ and $T_e$ are the equilibrium magnetic field, electron density and temperature at the position of rational surface $r_j$ respectively).

Table 2. Basic equilibrium parameters

| $\hat{s}$ | $q$ | $\bar{\eta}_i$ | $\varepsilon_n$ | $R$ [m] | $a$ [m] | $B$ [T] | $\tau_e$ | $n$ | $r_j$ [cm] | $T_e$ [eV] | $N_e$ [$10^{19}\text{m}^{-3}$] |
|---|---|---|---|---|---|---|---|---|---|---|---|
| 2 | 4 | 4.3 | 0.12 | 1.7 | 0.6 | 1.8 | 1 | -60 | 54 | 140 | 1.2 |

Some ITG related parameters required for Reynolds stress are listed in Table 3 (where the parameters $\eta^2$, $\beta_1$ and $\beta_2$ are related to ITG mode structure [11], $k_*(x_*)$, $\sigma$ and $x_0$ are related to Reynolds stress structure and defined in Appendix C, $\omega$ in this table stands for the real frequency of ITG mode).

Table 3. Partial mode related parameters

| $\eta^2$ | $k_*(x_*)$ | $\beta_1$ | $\beta_2$ | $\sigma$ | $x_0$ |
|---|---|---|---|---|---|
| -4+2$i$ | -0.002-0.002$i$ | 0.21-0.002$i$ | 0.43-0.08$i$ | 0.65 | -12.9 |
| $\omega$ [kHz] | $I_m(x_0)$ | $a_\mu$ | $\omega_G$ [kHz] | $L_{T_e}$ [cm] | $D(\tau_e)$ |
| -270 | $5\times 10^{-6}$ | 3 | 15.6 | 6 | 1 |

Table 4. Five radial points near the peak position of static Reynold stress $r_{j,0}$ and the radial position of $x_{\pm\infty}$

| $x_{-\infty} \Leftrightarrow r_{j,-12}$ [cm] | $r_{j,-2}$ [cm] | $r_{j,-1}$ [cm] | $r_{j,0}$ [cm] | $r_{j,1}$ [cm] | $r_{j,2}$ [cm] | $x_{+\infty} \Leftrightarrow a$ [cm] |
|---|---|---|---|---|---|---|
| 47.1 | 54.1 | 54.8 | 55.5 | 56.2 | 56.9 | 60 |

## 4. The periodic bursting of GAM for single central rational surface

In this section ITG is assumed to be the micro-turbulence generating zonal flow as suggested by Appendix A. The numerical results are presented for single central rational surface only. The more realistic case with coupling between multiple central rational surfaces is presented in the next section.

The periodic bursting of GAM can be clearly seen in figure 2, and figure 3 at five distinct radial positions (defined in Table 4, in which the radial positions of $x_{\pm\infty}$ are also given). One can also see that the periodic bursting of GAM is highly correlated with the downward as well as upward zero-crossing of radial group velocity, and the typical period ~ 4ms, is within the experimentally observed range, *e.g.*, 2-5ms in ASDEX-U [1], 2-4ms in T-10 [2], 1-3ms in HL-2A [5,6]. The structure of temporal evolution is spatially-dependent, this is also consistent with experimental observations in the sense that no temporal pattern is recognized so far. One can also observe that the GAM amplitudes increase with higher level of TLFZF from figure 2, since zonal flow as a whole are driven by inhomogeneous Reynolds stress term, as seen from Eq.(D1). The downward (upward) zero-crossing results in stronger (weaker) and longer (shorter)-lasting GAM. The similar patterns shown in figure 3 are also reported in JFT-2M (figure 2(d) of [4]), JET (figure 7 of [9]), and EAST (figure 8 of [10]).

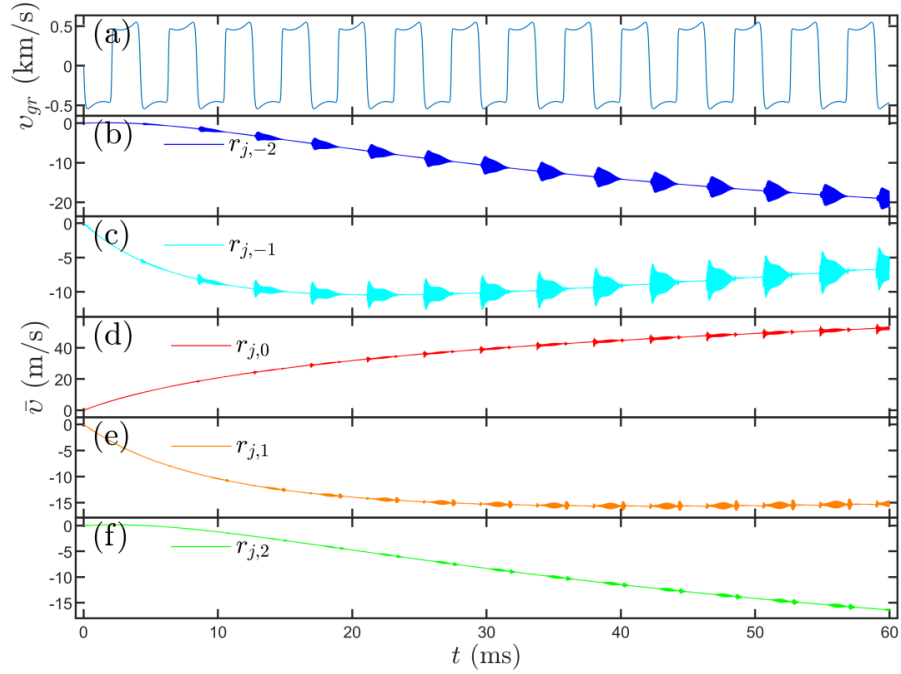

Figure 2. (a) Temporal evolution of radial group velocity and (b-f) zonal flow with intermittent excitation of GAM at five radial positions, (b) $r_{j,-2}$, (c) $r_{j,-1}$, (d) $r_{j,0}$, (e) $r_{j,1}$, (f) $r_{j,2}$

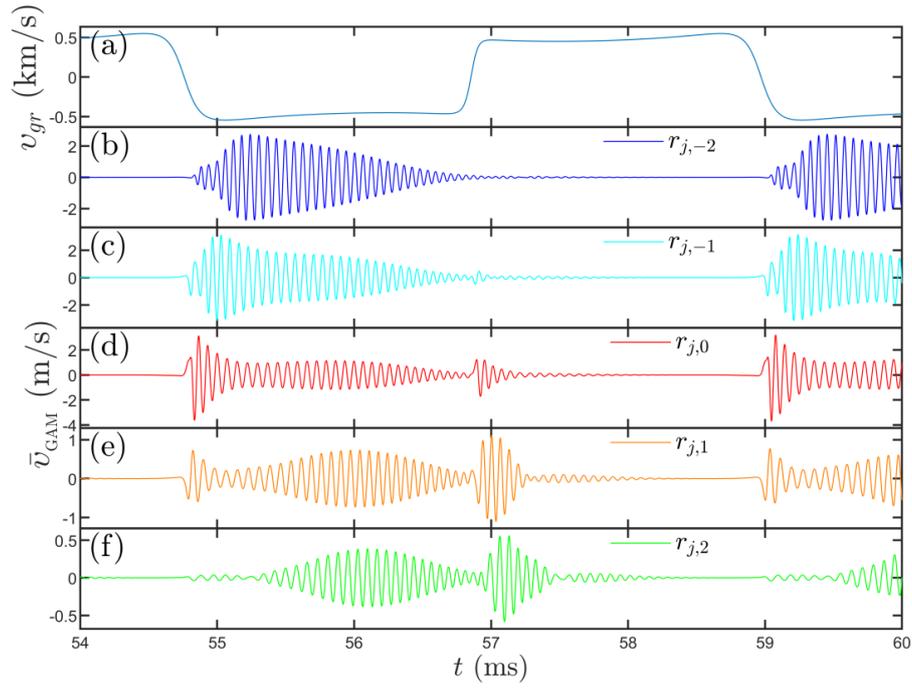

Figure 3. The magnified graphs in one period (54ms - 60ms) of figure 1 with low frequency portion filtered out

The frequency-time spectrogram, like that shown in figure 4(a), has been reported in ASDEX-U (figure 13(a) of [1]) and in T-10 (figure 20 of [2]) with similar bursting characteristics. It is also reported in HL-2A (figure 19 of [5] and figure 3(a) of [6]), however, with quite different bursting periods. Noticeably, the measurement in HL-2A is done very close to the plasma edge.

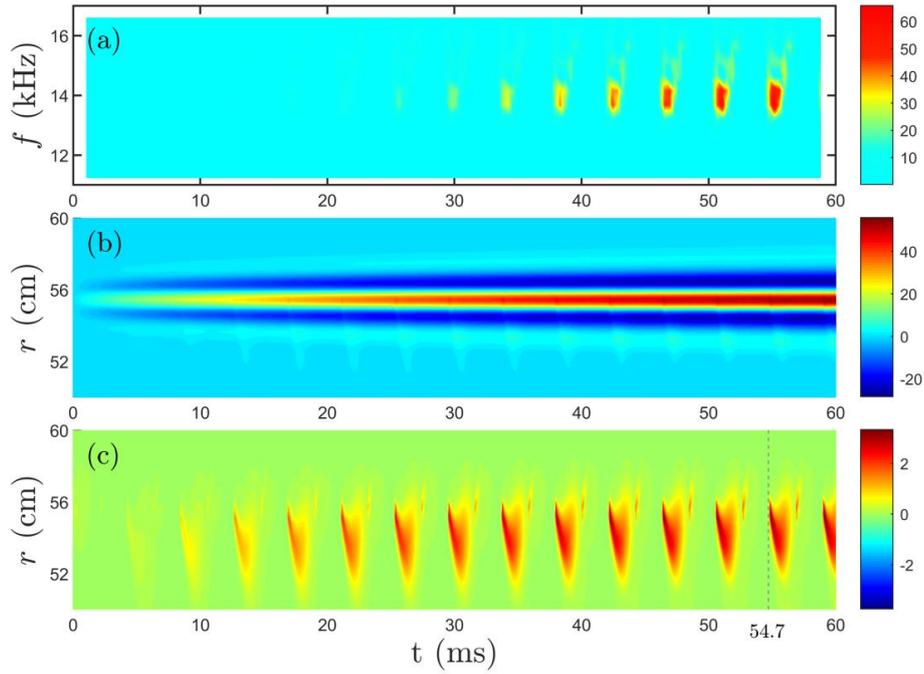

Figure 4. (a) Frequency-time spectrogram at $r_{j,0}$, (b) and (c) for spatiotemporal evolution for the zonal flow $\bar{\upsilon}$ and its high frequency component $\bar{\upsilon}_{GAM}$ respectively

The temporal evolution of spatial structure for the zonal flow $\bar{\upsilon}$ and its high frequency component $\bar{\upsilon}_{GAM}$ is presented in figure 4(b), (c) respectively. Similar spatiotemporal patterns, however, have not yet been reported in experiments - perhaps because of its fine radial structure, and inadequate spatial resolution of diagnostics. Right after 54.7ms, the so-called pre-GAM (around GAM frequency, but irregular spatial structure) is generated moving inwardly as driven by the ingoing instantons (see corresponding wave energy pattern in the movie in the caption of figure 6 (Multimedia view), and also in snapshots in figure 6(a), (b) and (c)). When the wave front hits the turning points, only a portion is reflected back outwardly. The coherent pattern is thus formed and evolves into a semi-eigenmode till ~ 56.83ms when the caviton decays into outgoing instantons. A shorter life-time pre-GAM is generated, but it gradually dissipates away. The reflection boundary at plasma edge has little effect on the subsequent behavior because the incoming wave near plasma edge is not strong enough to support a reflected wave that could reach the turning point to yield what could be classed as a full eigenmode [23].

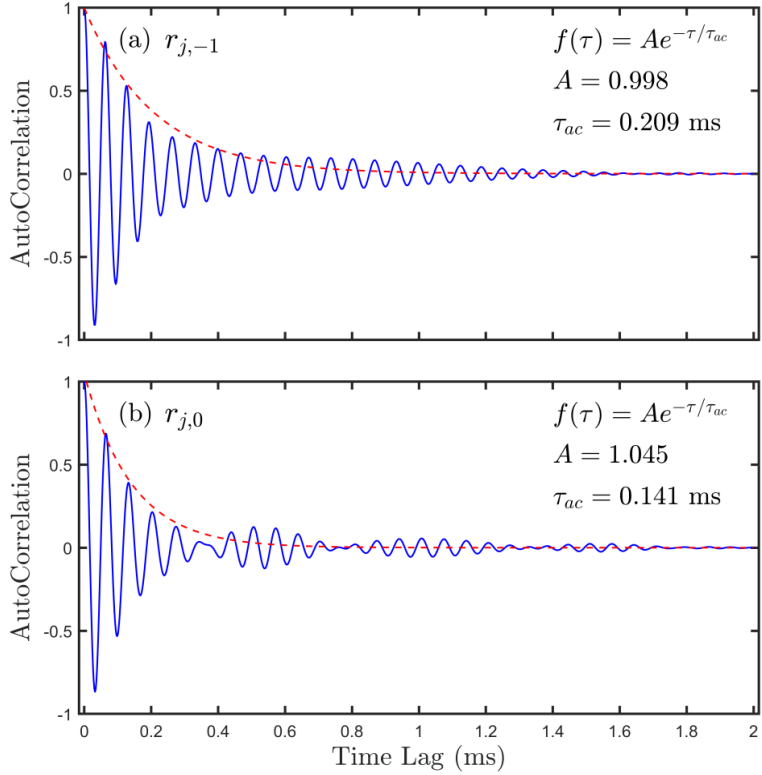

Figure 5. Auto-correlation function for different radial position at (a) $r_{j,-1}$ and (b) $r_{j,0}$

The quantitative study of GAM randomness has been documented in [7] by making use of auto-correlation function. The auto-correlation function defined by Eq.(B.2) in Appendix B of [7] is a very useful concept in experiment. It could reveal the nature of the observed intermittency, *i.e.* whether essentially stochastic or deterministic. The measurement is displayed in figure 15(a) and (b) of [7] for two different radial positions, also reported in figure 9 of [10]. Similar 'numerical measurements' are carried out and presented in figure 5 for two spatial positions. While the patterns are somewhat different from those in experiments they share one very important feature, a long tail after *non-exponential* quick fall-off. The non-exponential quick fall-off is inconsistent with $1/f$ scaling [1]; the long tail indicates long lasting dynamics in the system.

It is very important to note that the experimental data used for autocorrelation are filtered near GAM frequency. The temporal evolution of the perpendicular velocity is also reported using filtered data in Fig.8 of [10]. Both the reported figures are very similar to figure 5 presented here.

In addition to figure 4(c) the spatiotemporal evolution of GAM can, equivalently, be represented by the movie - 'GAM spatiotemporal evolution' in the caption of figure 6 (Multimedia view). Not only it captures the clear radial structure of GAM as a snapshot at any time, but also captures the temporal correlation to the motion of drift wave energy. Since the movie can only be viewed online, nine distinct snapshots are selected and displayed in figure 6. Two physical quantities, the high frequency zonal flow $\bar{\upsilon}_{\text{GAM}}$ and the normalized wave energy represented by $\cos^2\Theta$, are displayed jointly on the same time base during a cycle; the temporal (spatial) range is 54-60ms (48-60cm). This choice of spatial dimension is reasonable because: (1) the plasma edge is at 60cm, set to be the right boundary, *i.e.*, the reflection boundary for outgoing zonal flow; (2) the so-called turning point lies within 52-55cm as seen from figure 4(c), also from close-up figure 10(b) $\rho$ within 0.87-0.92; the latter serves to be the reflection layer for the ingoing pre-GAM. Let us see how the entire dynamics unfolds.

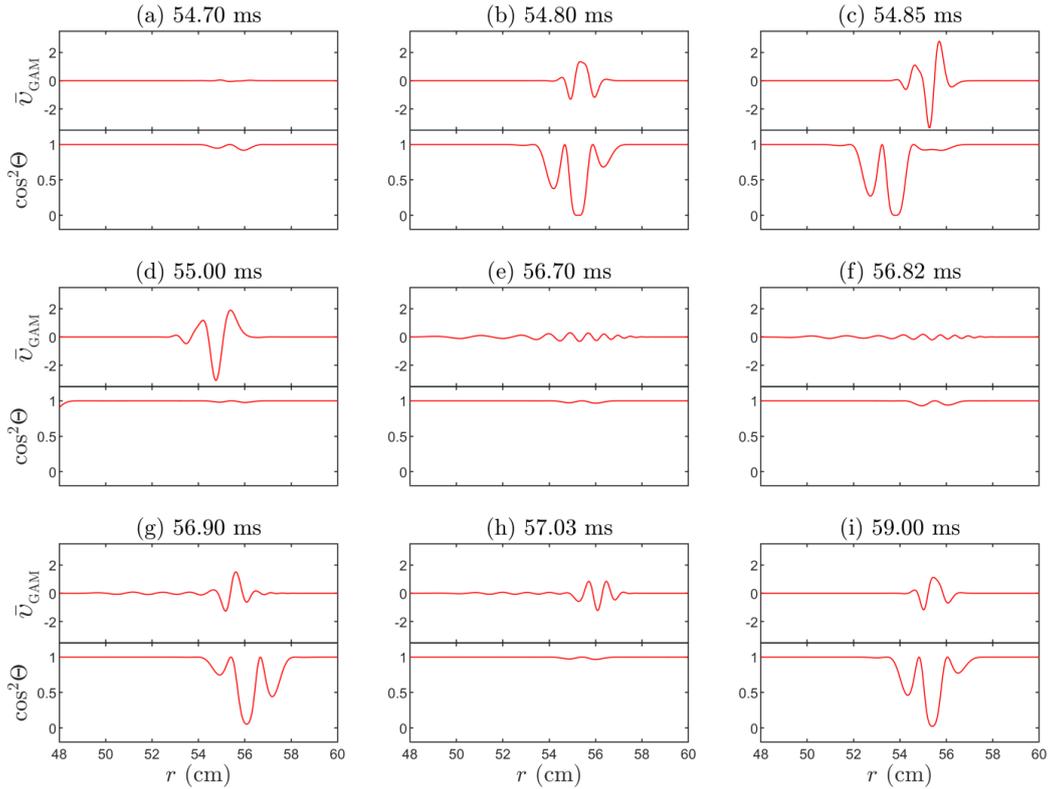

Figure 6. Snapshots of 9 time points in the movie, (a)54.70ms, (b)54.80ms, (c)54.85ms, (d)55.00ms, (e)56.70ms, (f)56.82ms, (g)56.90ms, (h)57.03ms, (i)59.00ms. The time evolution for 54-60ms can be seen via the link[1] 'GAM spatiotemporal evolution' (Multimedia view)

Initially, between 54-54.7ms, a pair of slowly breathing cavitons emerge, and begin to grow (figure 6(a)). During a very short period of time 0.1ms (from 54.7 to 54.8ms) three events occur, almost simultaneously: (1) The amplitude of normalized wave energy grows – crosses half and eventually becomes unity, (2) the pre-GAM starts to form rapidly in the reaction region (where static Reynolds stress is not small), and (3) the caviton pair decays into instantons (figure 6(b)). Right after 54.85ms instantons quickly propagate inwards and bring up the ingoing pre-GAM (figure 6(c)). The ingoing instantons disappear after 55ms and a new caviton pair starts to form and breathe slowly in the reaction region (figure 6(d)). At this moment, the pre-GAM front reaches the turning point; the penetrated part is then absorbed somewhere further inward, while the reflected part moves outwards. After that, the interference between the ingoing and outgoing zonal flow around GAM frequency results in the transit phase of forming an "eigenmode" between the turning point and the right edge of reaction region. At 56.7ms (figure 6(e)), the outgoing GAM reaches the right zero boundary (GAM cannot propagate outside the last closed magnetic surface), and then is reflected back to form semi-eigenmode between plasma edge and the turning point till 56.82ms (figure 6(f)). At this moment a caviton pair starts first grows and then decays into outgoing instantons. Right after 56.9ms, the instantons quickly propagate outwards (figure 6(g)) inducing right moving pre-GAM overlapped with the existing outgoing GAM. Such an interference results in a rather complicated pattern till 57.03ms (figure 6(h)), at that moment the outgoing instantons run outside the reaction region. Afterwards, the GAM gradually dissipated away. The process mentioned above occurs almost repetitively at 59ms when another caviton pair grows and decays into ingoing instantons (figure 6(i)). A new instance of pre-GAM is generated and moves inwardly just at that moment.

---

[1] http://home.ustc.edu.cn/~lzy0928/GAM%20spatiotemporal%20evolution.mp4

## 5. GAM in the coupling between multiple central rational surfaces

In previous sections the physics of GAM associated with a single rational surface is discussed. This is surely an idealized situation since coupling with nearby rational surfaces is bound to affect the GAM characteristics. It is particularly true in our case, since the width of reaction region pertaining to one central rational surface covers at least 11 neighbouring central rational surfaces (within a range of 1cm). Because the phase between the neighbouring rational surfaces is not definitive, the coupling would inevitably bring complexity into the full system, which is likely to be different from shot to shot, *i.e.*, irreproducibility will be automatic.

The zonal flow equation, in which coupling among 11 rational surfaces ($j_{max} = 5$) is included, comes out to be

$$\frac{\partial V}{\partial \tau} - \frac{1}{R_z}\frac{\partial^2 V}{\partial x^2} + \gamma_F V = -\psi q^2 \hat{\varpi}_G^2 \chi - \frac{\partial}{\partial x}\left\{\sum_{j=-j_{max}}^{j_{max}} \left(R_j[1] + \varepsilon R_j[\cos\vartheta]\right)\cos^2\Theta_j\right\}, \quad (25)$$

where $\gamma_F$ denotes the flow damping rate and is set to be a constant frequency (10Hz). The phase function associated with each rational surface, influenced by the same zonal flow, is

$$\Theta_j(x,\tau) = \int_0^\tau d\tau' V\left(x + j - \hat{s}\int_{\tau'}^\tau d\xi \hat{v}_{gr}\left(\vartheta_j + \vartheta(\xi)\right), \tau'\right), \quad (26)$$

where $\vartheta_j$ is the integral constant of the poloidal characteristic line, $d\vartheta(\xi)/d\xi = \hat{v}_{gy}(\vartheta(\xi))/(k_\vartheta r_j)$. It is reasonable to assume that $\vartheta_j$ is an arbitrary number in $[0, 2\pi]$, such that for each rational surface the zero crossing time of radial group velocity is different, this will imply shot to shot irregularity in GAM excitation. The equation for the first harmonic sinusoidal component of sound wave is

$$\begin{aligned}\left[\frac{\partial^2}{\partial \tau^2}\left(1 + D(\tau_e)\delta^2\frac{\partial^2}{\partial x^2}\right) + \psi\hat{\omega}_G^2 + (1+\tau_e)\frac{\delta^2}{R_z}\frac{\partial}{\partial \tau}\frac{\partial^4}{\partial x^4}\right]\chi = \\ \frac{2}{R_z}\frac{\partial^2 V}{\partial x^2} - 2\frac{\partial}{\partial x}\left\{\sum_{j=-j_{max}}^{j_{max}}\left(R_j[1] + \varepsilon R_j[\cos\vartheta]\right)\cos^2\Theta_j\right\} \\ + 2c_R\frac{\partial}{\partial \tau}\frac{\partial^2}{\partial x^2}\left\{\sum_{j=-j_{max}}^{j_{max}}\left(R_j[\sin\vartheta] + \frac{\varepsilon}{3}R_j[\sin 2\vartheta]\right)\cos^2\Theta_j\right\}\end{aligned} \quad (27)$$

Eqs.(25-27) are numerically solved for the same parameters as in section 3, except $I_m(x_0) = 5\times 10^{-5}$. Temporal evolution of zonal flow at five distinct radial positions are displayed in figure 7, its high frequency components are further magnified in the interval 54-60ms in figure 8. Notice that the periodic bursting observed in the single central rational surface treatment in section 4, disappears. What appears, instead, looks more like the intermittent excitation observed in tokamaks [3,4,8-10]. Figure 7(a) and figure 8(a) illustrate radial group velocity of five rational surfaces: $j = \pm 4$, $j = \pm 2$ and $j = 0$. Due to the arbitrary initial phase $\vartheta_j$, their zero-crossing time are different. As a result, GAMs are triggered asynchronously for different rational surfaces, then propagate radially and overlap one another, resulting in a rather intricate pattern. The frequency-time spectrogram of GAM at $r_{j,0}$ is displayed in figure 9, with similar intermittent characteristics as experimental results [1,2,5,6].

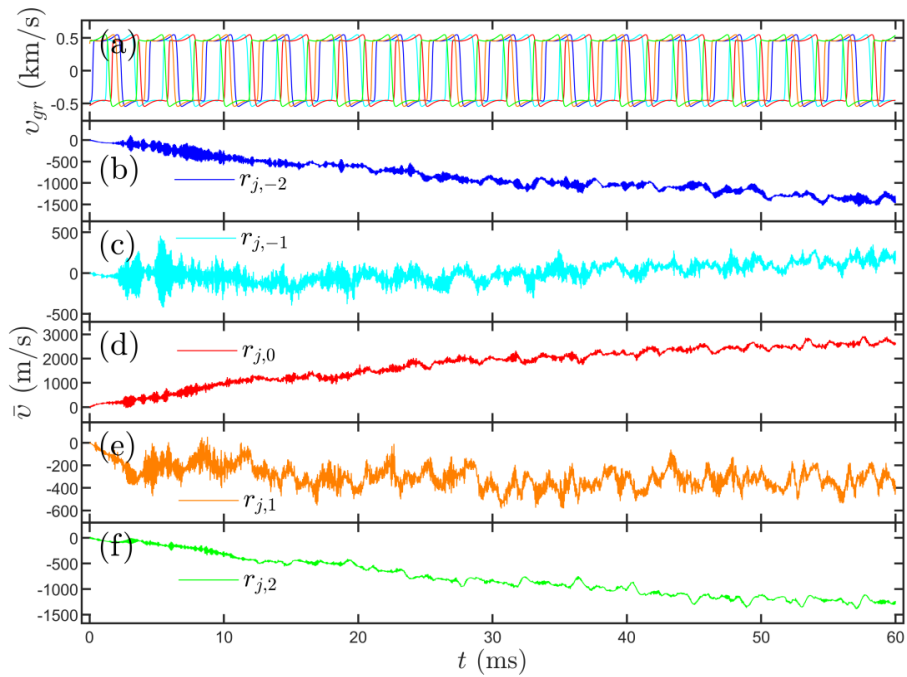

Figure 7. (a) Temporal evolution of radial group velocity of five rational surfaces: $j = -4$ (blue), $j = -2$ (cyan), $j = 0$ (red), $j = 2$ (crown), $j = 4$ (green), (b-f) zonal flow in the coupling between 11 rational surfaces at five radial positions, (b) $r_{j,-2}$, (c) $r_{j,-1}$, (d) $r_{j,0}$, (e) $r_{j,1}$, (f) $r_{j,2}$

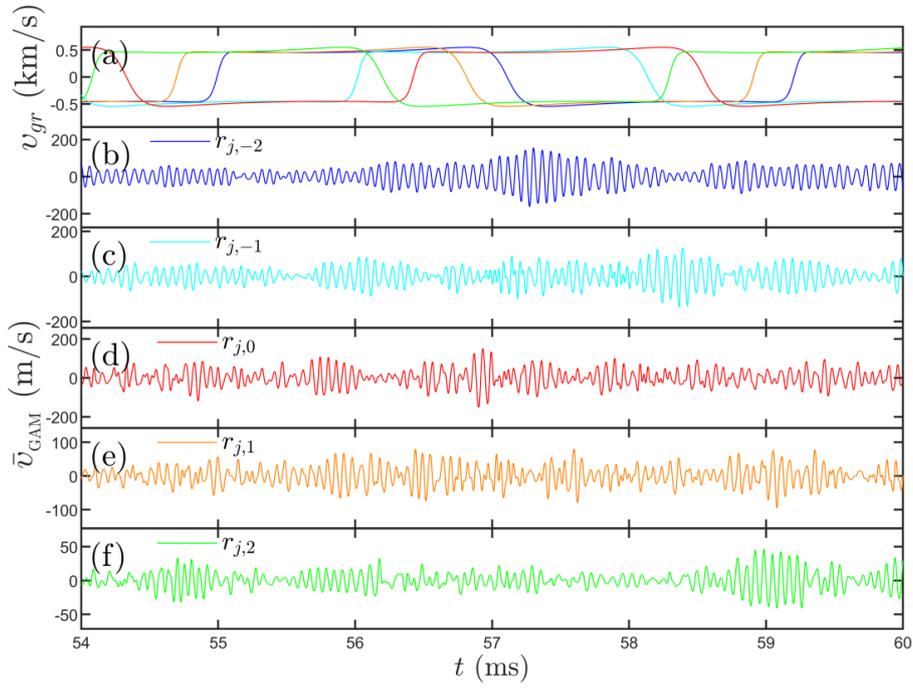

Figure 8. The magnified graphs in 54-60ms of figure 7 with low frequency portion filtered out

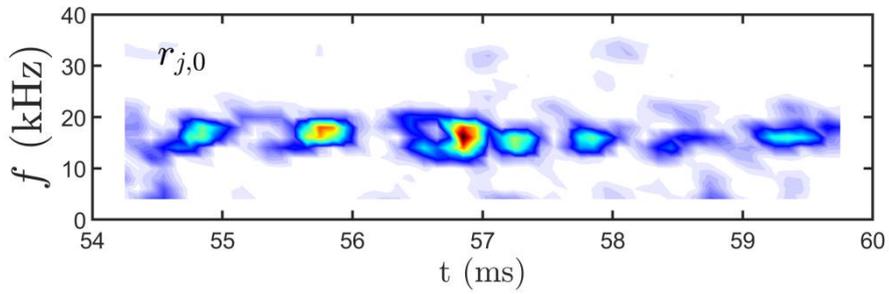

Figure 9. Frequency-time spectrogram at $r_{j,0}$

## 6. Impact of sign of weak dispersive coefficient on the GAM propagation

The GAM propagation is a hot topic since the first observation in experiments [21,22] till recently. The observed outward phase velocity seems consistent with the eigenmode theory between edge and WKB turning point in downward temperature profile [23,24]. The corresponding positive dispersive coefficient can only be derived by kinetic model [16,25,26] or by fluid model with FLR correction [27], not by fluid model without FLR correction [28,29]. On the other hand, the simulations in [28,29] emphasize the significance of group velocity, rather than phase velocity, because the former is the real physical entity. At this point we would like to point out that the propagation issue is not fully determined by property of media, but also by the way of GAM generation.

The eigenmode solution is the long-time asymptotic solution. In the picture developed in this paper, GAM is generated by instantons. The spatiotemporal correlation of instantons and GAM is vividly evident in figures 10-13(a) and (b). The Instantons run away rapidly. The initial value problem reveals the motion of the GAM so generated. The lifetime of GAM is not long enough to allow eigenmode formation. This may shed light on why GAM propagation may be observed in the simulation of [28,29]. It is fruitful, then, to investigate GAM propagation $\bar{\upsilon}_{GAM}$ under intermittency environment when there is a single central rational surface. The presentation is organized as follows: 1) The sub-section 6.1 (6.2) is for a positive (negative) dispersive coefficient, and 2) Each one consists of two groups, the first (second) being triggered by inward (outward) moving instantons. The dynamics of each group is illustrated through four 3D close-up figures: (a) motion of instantons, (b) spatiotemporal structure of $\bar{\upsilon}_{GAM}$, (c) spectrum versus frequency and spatial position, (d) spectrum versus wave number and time. A clarification of nomenclature is in order. The WKB turning point refers solely to eigenvalue problem. For wave propagation the inner (outer) turning point refers to the location of wave reflection where the group velocity reverse sign from inward (outward) to outward (inward).

## 6.1. $D(\tau_e) = 1$

The data have been presented in figure 4(c).

1) The ingoing GAM (being triggered by inward moving instantons, see figure 6(b),(c) and figure 10(a)) in period 54.6-56.8ms and $\rho = 0.85 - 0.97$ is displayed in figure 10(b-d). Initially both phase velocity and the group velocity are inward ($\upsilon_p < 0, \upsilon_g < 0$) till hitting the turning point around 55.5 ms. After being reflected, GAM moves outwardly with $\upsilon_p > 0$ and $\upsilon_g > 0$. This is consistent with gyrokinetic simulations [35-41] and fluid model with FLR correction [42].

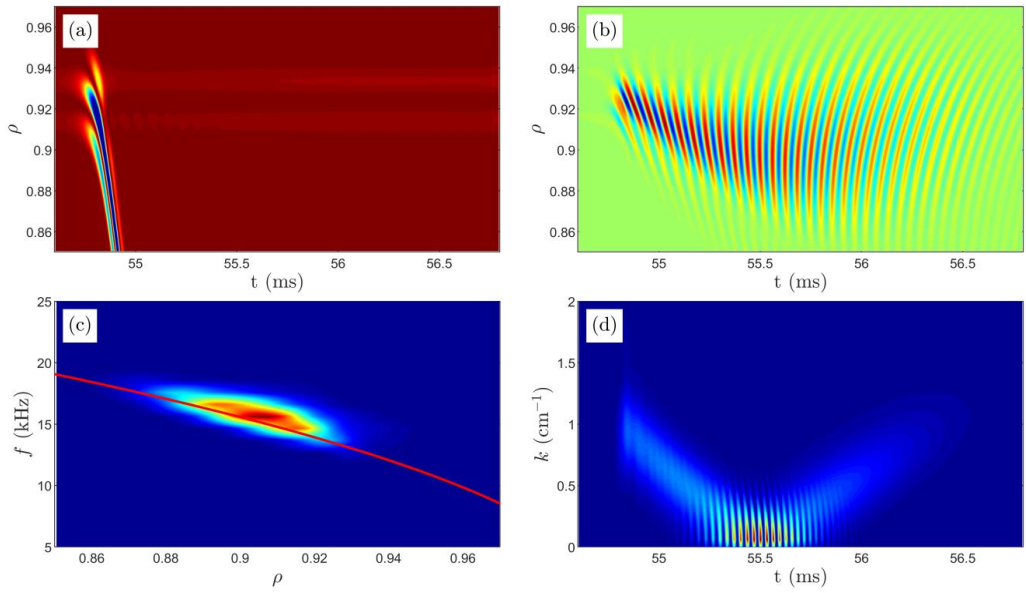

Figure 10. Spatiotemporal structure of (a) drift wave energy $I_m \propto \cos^2 \Theta$ and (b) $\bar{\upsilon}_{GAM}$ for $D(\tau_e) = 1$ in period 54.6-56.8ms and $\rho = 0.85 - 0.97$, (c) spectrum versus frequency and spatial position, (d) spectrum versus wavenumber and time

2) The outgoing GAM (being triggered by outward moving instantons, see figure 6(f),(g) and figure 11(a)) in period 56.8-57.4ms and $\rho = 0.91 - 0.96$ is displayed in figure 11(b-d). Initially both the phase and group velocity of GAM are positive, $\upsilon_p > 0$ and $\upsilon_g > 0$, which is consistent with gyrokinetic simulation [43,44].

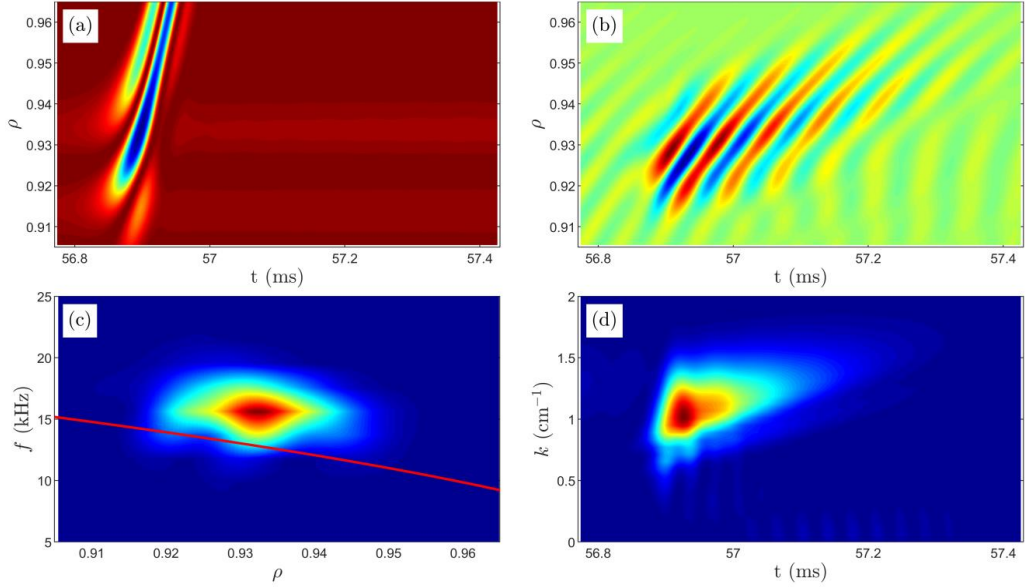

Figure 11. Spatiotemporal structure of (a) drift wave energy $I_m \propto \cos^2 \Theta$ and (b) $\bar{v}_{\text{GAM}}$ for $D(\tau_e) = 1$ in period 56.8-57.4ms and $\rho = 0.91 - 0.96$, (c) spectrum versus frequency and spatial position, (d) spectrum versus wavenumber and time

## 6.2. $D(\tau_e) = -1$

Except for the change of $D(\tau_e)$ from 1 to -1, all other parameters remain intact. They are displayed here in order to compare with simulations using fluid model without FLR correction. The intermittent characteristics of GAM are almost the same as results in figures 2-5. The only difference is the propagation behavior of GAM, phase velocity in opposite direction to group velocity, in contrast to same direction for $D(\tau_e) > 0$. The outward propagating GAM does exist in a negative dispersive media ($D(\tau_e) < 0$); but brought by inward instantons. In a sense the 'emergence' of GAM in simulation for negative $D(\tau_e)$ has the special theoretical significance. The formation of eigenmode requires two WKB turning points. The linear differential equation is unable to get inner WKB turning point in this case.

1) The activity is triggered by inward moving instantons (in figure 12(a)). Spatiotemporal

structure of GAM in period 54.6-56.8ms and $\rho = 0.88 - 0.96$ is displayed in figure 12(b-d). Initially the inward moving instantons (in figure 12(a)) trigger inward phase velocity ($\upsilon_p < 0$), but outward group velocity ($\upsilon_g > 0$), which is consistent with Landau fluid simulation [28,29]. After being reflected at the outer turning point, GAM moves inwardly with $\upsilon_p > 0$ and $\upsilon_g < 0$. This behavior has not yet been reported in [28,29], perhaps the signal is too weak to be noticed. Spectrum in figure 12(c) is very similar to figure 10(c) for almost same frequency at turning point and local $\omega_G$.

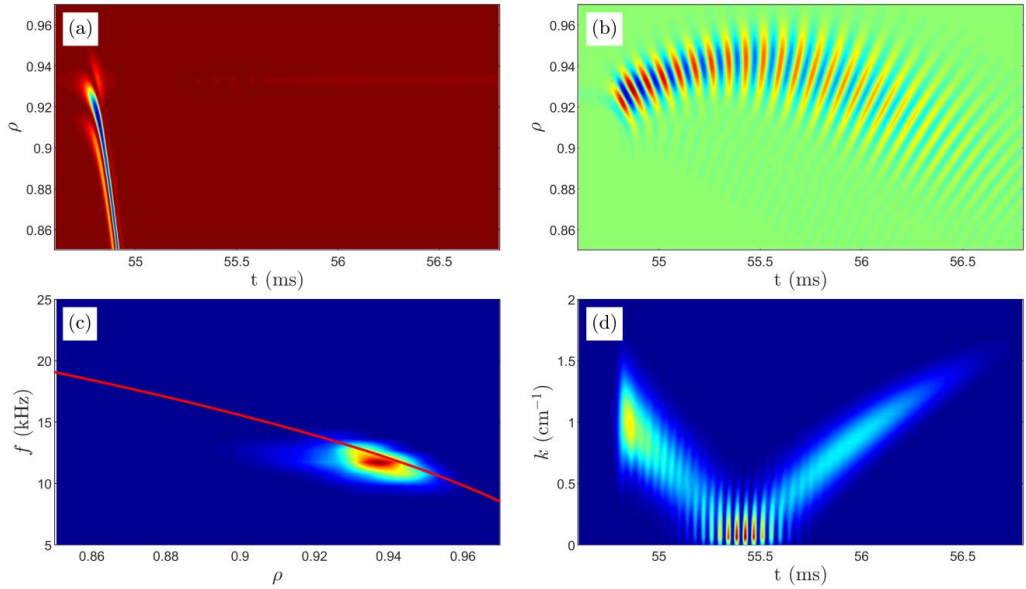

Figure 12. Spatiotemporal structure of (a) drift wave energy $I_m \propto \cos^2 \Theta$ and (b) $\bar{\upsilon}_{\text{GAM}}$ for $D(\tau_e) = -1$ in period 54.6-56.8ms and $\rho = 0.86 - 0.96$, (c) spectrum versus frequency and spatial position, (d) spectrum versus wavenumber and time

2) The activity is triggered by outward moving instantons (in figure 13(a)). Spatiotemporal structure of GAM is displayed in figure 13(b) for 56.8-57.7ms and $\rho = 0.86 - 0.96$, the phase and group velocity of GAM are opposite, *i.e.*, $\upsilon_p > 0$ and $\upsilon_g < 0$. The frequency of GAM is a little smaller than the local $\omega_G$ and does not change as it propagates inwardly as shown in figure 13(c). Spectrum in figure 13(d) is very similar to figure 11(d).

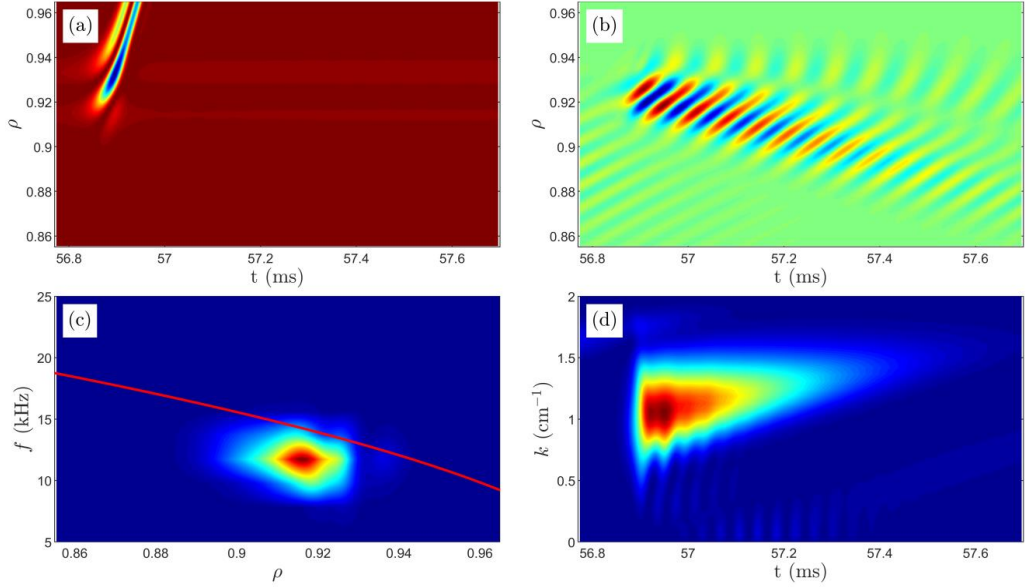

Figure 13. Spatiotemporal structure of (a) drift wave energy $I_m \propto \cos^2 \Theta$ and (b) $\bar{\upsilon}_{\text{GAM}}$ for $D(\tau_e) = -1$ in period 56.8-57.7ms and $\rho = 0.86 - 0.96$, (c) spectrum versus frequency and spatial position, (d) spectrum versus wavenumber and time

In summary, for $D(\tau_e) = 1$ the data are consistent with simulations based on gyrokinetic model and fluid model with FLR correction; for $D(\tau_e) = -1$ the data in figure 12(b) before 55.5ms are consistent with results in [28,29]; no similar data are reported in [28,29] for either in figure 12(b) after 55.5ms and in figure 13(b).

## 7. Conclusions

In this paper the theoretical model for toroidal zonal flow, including both TLFZF and GAM, is systematically developed based on Braginskii model with FLR correction (to weak dispersive coefficient) in tokamak configuration. This model is numerically solved by making use of the ITG mode structure in real configuration (figure 1) on basis of [11] to calculate two important meso-scale physical quantities: Reynolds stress and group velocity of ITG. It reveals four new high-level qualitative features regarding intermittent excitation and propagation of GAM.

(1) The numerical experiment based on the deterministic toroidal zonal flow system, Eq.(25-27) for multiple central rational surface coupling, seems to reproduce many intermittent characteristics of GAM observed in experiments in a variety of tokamaks [1-10]. For single central rational surface, Eqs.(18-20), however, the intermittent excitation reduces to a series of periodic bursting. The indefinite initial phase difference among multiple central rational surfaces is the likely cause behind the intermittent characteristics.

(2) The GAM generation is triggered when the radial group velocity of the drift wave crosses zero; it is the precise moment for the phase transition from a pair of caviton into instantons. The pulses brought by rapid motion of instantons induce resonance to the GAM dispersion. This can be better described in views of single central rational surface. The instantons immediately run away and periodic stopping follows as GAM is dissipated until the next occurrence of zero crossing.

(3) Not only GAM generation, but also GAM propagation is highly correlated with the (inward/outward) motion of instantons. Notice that the GAM propagation phenomenon is the solution of an initial value problem before the system has time to settle into an eigenstate. The initial condition is provided by the motion of triggering instantons. The GAM (inward/outward) propagation is systematically explored for both positive and negative dispersive coefficient. The results are compared with simulations in both gyrokinetic and fluid model in literature.

(4) We have also noticed that regardless the motion of triggering instantons, either inward or outward, the GAM phase-group velocity is always in the same (opposite) direction for positive (negative) dispersive media $D(\tau_e) = 1$ ($D(\tau_e) = -1$). In order to determine the sign of dispersive coefficient experimentally, one may have to measure both the phase velocity and the group velocity of GAM simultaneously. Either sign of either velocity may occur for both dispersive coefficients.

Finally, we must apprize the reader of the somewhat limited scope of this paper. Many interesting GAM phenomena observed in experiments - such as 'continuum' versus 'eigenmode', coexistence of multiple modes, radial extension of 'eigenmode' and the scaling law *etc.* - are not discussed here. We have concentrated here (through the numerical solutions of the three equations) only on a subset of GAM related issues. A more complete and systematic exploration (including for example working out the parameter dependences) is deferred to a future paper. Furthermore, an invariant turbulence level was assumed to calculate the time evolution of the system. Since the generated shear flow tends to suppress turbulence, this assumption is not fully valid - we expect quantitative modifications to our results through shear suppression. However, the theory developed in [11] as well as in this paper, limited to the like-mode coupling either in calculation of Reynolds stress and concepts regarding drift wave energy, will give a qualitative knowledge of the dynamics. Unlike-mode coupling is surely possible and it may lead to new features - a resonant mechanism [45], for instance.

**Appendix A Conditions for detrapping due to collision and occurrence of ITG in L-mode discharge near tokamak edge**

There is circulating a belief that GAM is likely generated by TEM not by ITG mode, because at the tokamak edge $\eta_i$ is normally less than 2, the threshold for instability.

In fact, in L-mode discharges near tokamak edge, the collision frequency is high enough that particle's trapping is minimal. Mathematically, this condition translates into $\omega_b \tau_{\text{detrap}} < 1$, where the bounce frequency $\omega_b = (\upsilon_{T_e}/qR)(r/R)^{1/2}$ and the de-trapping time $\tau_{\text{detrap}} \simeq (r/R)/\nu_{ei}$ ($\upsilon_{T_e} \equiv \sqrt{T_e/m_e}$ is the electron thermal speed and $\nu_{ei} \sim N_e Z_{eff}^2 / T_e^{3/2}$ is the electron-ion collision frequency, $Z_{eff}$ is the effective ion charge $Z$). For typical tokamak geometry and plasma densities, the critical temperature, below which particle trapping is a minor effect, is a few hundred eV [46]. Experimental Data on a variety of tokamaks studying zonal flows is shown in Table 5 (with $Z_{eff} = 1$). Since $\omega_b \tau_{\text{detrap}} < 1$ is true for all cases, the trapped density will most likely be too small to support a TEM.

Table 5. Equilibrium parameters of L-mode (Ohmic) discharge in the edge ($r/a = 0.9$) of 7 tokamak machines

| Tokamaks | Shots | $q$ | $N_e$ [$\times 10^{19}$m$^{-3}$] | $T_e$ [eV] | $\nu_{ei}$ [kHz] | $\omega_b$ [kHz] | $\omega_b \tau_{\text{detrap}}$ |
|---|---|---|---|---|---|---|---|
| ASDEX [47] | #20787 (Ohmic) | 3.0 | 1.3 | 180 | 227 | 595 | 0.72 |
| T-10 [2] | #36819 (L-mode) | 3.5 | 1.0 | 200 | 151 | 480 | 0.57 |
| JFT-2M [3,4] | (L-mode) | 2.8 | 1.0 | 170 | 190 | 687 | 0.75 |
| HL-2A [5,6] | (L-mode) | 3.0 | 0.65 | 96 | 285 | 390 | 0.30 |
| DIII-D [7] | #141958 (L-mode) | 4.0 | 1.2 | 140 | 300 | 412 | 0.44 |
| DIII-D [7] | #142121 (L-mode) | 4.0 | 2.6 | 220 | 332 | 517 | 0.49 |
| JET [48,49] | #86470 (Ohmic) | 3.2 | 2.0 | 300 | 165 | 424 | 0.78 |
| JET [48,49] | #90492 (L-mode) | 3.0 | 2.7 | 300 | 220 | 452 | 0.62 |
| EAST [10] | #74036 (L-mode) | 4.9 | 1.1 | 100 | 447 | 212 | 0.10 |

The case for the ITG turbulence is a little stronger; there are published papers measuring GAMs in which the edge ($\rho \equiv r/a \approx 0.9$) conditions are such [$\eta_i = L_n / L_{T_i} > 2$] that unstable ITG can readily exist. Table 6 lists edge-conditions for four machines; the data sources are: figure 5 of [47] for ASDEX, figure 3 of [7] for DIII-D, figure 15 of [9] for JET, and figure 3 of [10] for EAST. All shots, except for #18813 that provides ion temperature profile, are analyzed by assuming $L_{T_i} = L_{T_e}$.

Table 6. $\eta_i$ in L-mode (Ohmic) discharge near tokamak edge ($\rho \equiv r/a \approx 0.9$) on 4 machines

| Tokamaks | Shots | $L_n$ [cm] | $L_T$ [cm] | $\eta_i$ |
|---|---|---|---|---|
| ASDEX [47] | #20787 (Ohmic) | 9.7 | 4.3 | 2.2 |
| ASDEX [47] | #18813 (Ohmic) | 18.7 | 7.4 | 2.5 |
| DIII-D [7] | #141958 (L-mode) | 20.1 | 5.9 | 3.4 |
| DIII-D [7] | #142121 (L-mode) | 37.8 | 10.2 | 3.7 |
| JET [8,9] | #87802 (Ohmic) | 22.3 | 7.3 | 3.1 |
| EAST [10] | #74036 (L-mode) | 8.2 | 3.8 | 2.2 |

## Appendix B Derivation of charge and particle conservation equation for axisymmetric electrostatic mode in meso-scale

In this Appendix the derivation of charge and particle conservation for low frequency electrostatic axisymmetric mode, Eqs.(1), (2) respectively, is presented on basis of Braginskii two-fluid model [17] essentially same as [18], but extended from cold ion to warm ion.

It begins with symbol definition. The quantities having both over-bar and under-tilde stand for fluctuations in meso-scale, while those without overbar and tilde stand for equilibrium with subscript $i$ ($e$) representing ion (electron). Those quantities having overtilde in lower case stand for fluctuating field in microcopic scale like $\tilde{\phi}$. $\boldsymbol{u}$, $\boldsymbol{J}$, $N$, $P$ and $\varphi$ are fluid velocity, current, density, pressure and potential respectively, $\mathbf{\Pi}$ is viscosity tensor, $\mathbf{R}$ is friction force ($\mathbf{R}_i = -\mathbf{R}_e$), $\mathbf{B}$ is magnetic field, $\mathbf{b}$ is the unit vector along the equilibrium magnetic field, $T_{i,e}$ is ion (electron) temperature, $c$ is the speed of light, and $e$ is the unit electric charge.

The ion momentum equation in meso-scale is obtained by ensemble averaging in micro-scale over the full ion momentum equation containing both micro- and meso- scale [18], provided that the scale separation is valid.

$$m_i N \left( \frac{\partial \bar{\mathbf{u}}_i}{\partial t} + \bar{\mathbf{u}}_i \cdot \nabla \bar{\mathbf{u}}_i + \langle \tilde{\mathbf{u}}_E \cdot \nabla \tilde{\mathbf{u}}_E \rangle_{en} \right) = eN \left( \bar{\mathbf{E}} + \frac{1}{c} \bar{\mathbf{u}}_i \times \mathbf{B} \right) - \left( \nabla \bar{P}_i + \nabla \cdot \bar{\mathbf{\Pi}}_i + \bar{\mathbf{R}}_i \right), \quad \text{(B.1)}$$

where $\langle \tilde{\mathbf{u}}_E \cdot \nabla \tilde{\mathbf{u}}_E \rangle_{en} = \mathbf{b} \times \bar{\mathbf{u}}_{NP}$, $\tilde{\mathbf{u}}_E \equiv (c/B) \mathbf{b} \times \nabla \tilde{\phi}$, and $\bar{\mathbf{u}}_{NP} \equiv \langle (c^2/B^2)(\mathbf{b} \times \nabla \tilde{\phi} \cdot \nabla) \nabla_\perp \tilde{\phi} \rangle_{en}$ is called the nonlinear polarizaion drift velocity. This term represents the process that two microscale electrostatic potentials (like drift waves) $\tilde{\phi}$ are annihilated into one velocity field in mesoscale, known as 'three wave interaction' in literature. $\langle ... \rangle_{en}$ stands for ensemble average over microscopic scale, $i.e.$ $\bar{\mathbf{u}}_{NP}$ is in mesoscale.

To the leading order for low frequency waves $\omega \ll \omega_{ci}$, where $\omega_{ci} \equiv eB/cm_i$ is the ion cyclotron frequency, the perpendicular component of Eq.(B.1) is chosen to be the electrostatic and diamagnetic drift velocity

$$\bar{\mathbf{u}}_{i,\perp}^{(0)} = \bar{\mathbf{u}}_E + \bar{\mathbf{u}}_{i,D}, \quad \bar{\mathbf{u}}_E \equiv \frac{c}{B} \mathbf{b} \times \nabla \bar{\varphi}, \quad \bar{\mathbf{u}}_{i,D} \equiv \frac{c}{eNB} \mathbf{b} \times \nabla \bar{P}_i. \quad \text{(B.2)}$$

Neglecting friction force $\bar{\mathbf{R}}_i$, and substituting $\bar{\mathbf{u}}_{i,\perp}^{(0)}$ into the inertia term and viscosity tensor $\bar{\mathbf{\Pi}}_i$ yield

$$\bar{\mathbf{u}}_{i,\perp} = \bar{\mathbf{u}}_E + \bar{\mathbf{u}}_{i,D} + \frac{1}{\omega_{ci}} \mathbf{b} \times \left( \frac{\partial}{\partial t} + \bar{\mathbf{u}}_{i,\perp}^{(0)} \cdot \nabla \right) \bar{\mathbf{u}}_E + \frac{c}{eNB} \mathbf{b} \times \nabla \cdot \left( \bar{\mathbf{\Pi}}_{i,\parallel} + \bar{\mathbf{\Pi}}_{i,\perp} \right). \quad \text{(B.3)}$$

In Eq.(B.3) $\bar{\mathbf{\Pi}}_{i,\parallel}$ and $\bar{\mathbf{\Pi}}_{i,\perp}$ are parallel and perpendicular viscosity tensor respectively. Eq.(B.3) is achieved by making use of the so-called Hinton-Horton cancellation [50]

$$m_i N \left( \frac{\partial}{\partial t} + \bar{\boldsymbol{u}}_{i,\perp}^{(0)} \cdot \nabla \right) \bar{\boldsymbol{u}}_{i,D} + \nabla \cdot \bar{\bar{\boldsymbol{\Pi}}}_{i,g} = 0. \tag{B.4}$$

In Eq.(B.4) $\bar{\bar{\boldsymbol{\Pi}}}_{i,g}$ is the gyro-viscosity. For axisymmetric mode $\boldsymbol{b} \times \bar{\boldsymbol{u}}_{i,\perp}^{(0)} \cdot \nabla \bar{\boldsymbol{u}}_E$ can be neglected. Then, the expression for perpendicular velocity in terms of electrostatic potential and density fluctuation

$$\bar{\boldsymbol{u}}_{i,\perp} = \frac{c}{B} \boldsymbol{b} \times \left( \nabla \bar{\varphi} + \frac{T_i}{eN} \nabla \bar{N} \right) - \frac{1}{\omega_{ci}} \frac{c}{B} \frac{\partial}{\partial t} \nabla_\perp \bar{\varphi} + \frac{1}{Nm_i \omega_{ci}} \boldsymbol{b} \times \nabla \cdot \bar{\bar{\boldsymbol{\Pi}}}_{i,\perp} - \frac{1}{\omega_{ci}} \bar{\boldsymbol{u}}_{\mathrm{NP}}, \tag{B.5}$$

where $\bar{\bar{\boldsymbol{\Pi}}}_{i,\perp}$ is retained to avoid numerical divergence at short wavelength. The parallel viscosity $\bar{\bar{\boldsymbol{\Pi}}}_{i,\|}$, however, is neglected.

The electron momentum equation is

$$m_e N \left( \frac{\partial \bar{\boldsymbol{u}}_e}{\partial t} + \bar{\boldsymbol{u}}_e \cdot \nabla \bar{\boldsymbol{u}}_e + \langle \tilde{\boldsymbol{u}}_E \cdot \nabla \tilde{\boldsymbol{u}}_E \rangle_{\mathrm{en}} \right) = -eN \left( \bar{\boldsymbol{E}} + \frac{1}{c} \bar{\boldsymbol{u}}_e \times \boldsymbol{B} \right) - \left( \nabla \bar{P}_e + \nabla \cdot \bar{\bar{\boldsymbol{\Pi}}}_e + \bar{\boldsymbol{R}}_e \right). \tag{B.6}$$

The parallel Ohm law can be derived from the parallel component of Eq.(B.6) straightforwardly by invoking the concept of conductivity $\bar{R}_{e\|} = -eN\eta_\| \bar{J}_\|$, where $\bar{J}_\|$ is the parallel current fluctuation, $\eta_\| \equiv m_e \nu_{ei} / 2Ne^2$. Simply, neglecting the electron inertia and viscosity tensor yields

$$\eta_\| \bar{J}_\| = (\boldsymbol{b} \cdot \nabla) \left( \frac{T_e}{eN} \bar{N} - \bar{\varphi} \right). \tag{B.7}$$

In Eq.(B.7) use is made of the *assumed* state equation $\bar{P}_e := T_e \bar{N}$.

The one-fluid equation of motion can be derived by adding Eq.(B.1) to Eq.(B.6) with dropping electron inertia term and viscosity tensor

$$m_i N \left( \frac{\partial \bar{\boldsymbol{u}}}{\partial t} + \bar{\boldsymbol{u}} \cdot \nabla \bar{\boldsymbol{u}} + \boldsymbol{b} \times \bar{\boldsymbol{u}}_{\mathrm{NP}} \right) = -\left( \nabla \bar{P} + \nabla \cdot \bar{\bar{\boldsymbol{\Pi}}}_i \right) + \frac{1}{c} \bar{\boldsymbol{J}} \times \boldsymbol{B}, \quad \bar{P} := (T_e + T_i) \bar{N}. \tag{B.8}$$

In Eq.(B.8) $\bar{\boldsymbol{J}} \equiv eN(\bar{\boldsymbol{u}}_i - \bar{\boldsymbol{u}}_e)$, the subscript $i$ of $\bar{\boldsymbol{u}}_i$ has been omitted.

In the above manipulations the assumed state equation introduced cut-off to energy conservation. This makes sense provided that the thermal physics merely plays minor role for physics of zonal flow.

The two conservation equations, namely the particle conservation and charge conservation, are originally

$$\frac{\partial \bar{\tilde{N}}}{\partial t} + \bar{\tilde{\bm{u}}} \cdot \nabla N + N \nabla \cdot \bar{\tilde{\bm{u}}} = 0, \tag{B.9}$$

and

$$\nabla \cdot \bar{\tilde{\bm{J}}} = 0. \tag{B.10}$$

By taking the vector product of Eq.(B.8) with $\bm{B}$, we obtain

$$\bar{\tilde{\bm{J}}}_\perp = \frac{c}{B^2} \bm{B} \times \nabla \bar{\tilde{P}} - \frac{c^2 m_i N}{B^2} \frac{\partial}{\partial t} \nabla_\perp \bar{\tilde{\varphi}} + \frac{c}{B} \bm{b} \times \nabla \cdot \bar{\tilde{\bm{\Pi}}}_{i,\perp} - \frac{c m_i N}{B} \bar{\tilde{\bm{u}}}_{\text{NP}}. \tag{B.11}$$

In Eq.(B.11) the polarization drift in $\bar{\tilde{\bm{u}}}_\perp$ proportional to $(\omega/\omega_{ci})^2$ has been neglected.

Since the term

$$\frac{c}{B^2} \nabla \cdot (\bm{B} \times \nabla \bar{\tilde{P}}) = \frac{4\pi}{B^2 r} \left[ \frac{c}{B} \frac{dP}{dr} \left(1 - \frac{2r}{R} \cos\vartheta\right) + \langle E_\parallel / \eta_\parallel \rangle \frac{r}{qR} \right] \frac{d\bar{\tilde{P}}}{d\vartheta} \tag{B.12}$$

can be neglected for low-$\beta$ plasmas [51] ($\beta$ is the ratio of thermal pressure to magnetic pressure), the main contribution to divergence of diamagnetic current comes from the curvature of magnetic field in tokamak. However, the perpendicular viscosity $(c/B^2) \nabla \cdot (\bm{B} \times \nabla \cdot \bar{\tilde{\bm{\Pi}}}_{i,\perp})$ has to be retained for numerical computation. For the divergence of linear and nonlinear polarization current, curvature effect can be neglected because $|\bm{\kappa}| \ll |\nabla|$, where $|\bm{\kappa}|$ and $|\nabla|$ correspond to equilibrium scale and mesoscale respectively for $\bm{\kappa} := \nabla B / B$ is the curvature of magnetic field. Substituting the divergence of Eq.(B.7) and (B.11) into (B.10) yields charge conservation equation

$$\nabla \cdot \left[ \boldsymbol{b}(\boldsymbol{b} \cdot \nabla) \frac{1}{\eta_{\parallel}} \left( \frac{T_e}{eN} \bar{N} - \bar{\varphi} \right) \right] - (T_e + T_i) \frac{2c}{B} \boldsymbol{\kappa} \times \boldsymbol{b} \cdot \nabla \bar{N}$$
$$- \frac{c^2 m_i N}{B^2} \frac{\partial}{\partial t} \nabla_{\perp}^2 \bar{\varphi} + \frac{c}{B} \nabla \cdot \boldsymbol{b} \times \nabla \cdot \bar{\boldsymbol{\Pi}}_{i,\perp} - \frac{cm_i N}{B} \nabla \cdot \bar{\boldsymbol{u}}_{NP} = 0 \quad \text{(B.13)}$$

Similar procedure is adopted for obtaining the divergence of $\bar{\boldsymbol{u}}_{\perp}$. Substituting it into Eq.(B.9) yields particle conservation equation

$$\frac{1}{N} \frac{\partial \bar{N}}{\partial t} + \nabla \cdot (\bar{u}_{\parallel} \boldsymbol{b}) - \frac{2c}{B} \boldsymbol{\kappa} \times \boldsymbol{b} \cdot \nabla \left( \bar{\varphi} + \frac{T_i}{eN} \bar{N} \right) - \frac{1}{\omega_{ci}} \frac{c}{B} \frac{\partial}{\partial t} \nabla_{\perp}^2 \bar{\varphi}$$
$$+ \frac{1}{Nm_i \omega_{ci}} \nabla \cdot \boldsymbol{b} \times \nabla \cdot \bar{\boldsymbol{\Pi}}_{i,\perp} - \frac{1}{\omega_{ci}} \nabla \cdot \bar{\boldsymbol{u}}_{NP} = 0 \quad \text{(B.14)}$$

In Eq.(B.14) the term $\bar{\boldsymbol{u}} \cdot \nabla N$ is neglected, since for axisymmetric mode $\bar{\boldsymbol{u}}$ is mainly in poloidal direction.

The parallel component of Eq.(B.8) is

$$m_i N \partial_t \bar{u}_{\parallel} = -(T_e + T_i) \nabla_{\parallel} \bar{N}. \quad \text{(B.15)}$$

In terms of the dimensionless quantities, $\bar{N}/N \to \bar{n}$, $e\bar{\varphi}/T_e \to \bar{\varphi}$, and defining $\tau_i \equiv T_i/T_e$, $\rho_s \equiv c_s/\omega_{ci}$, $c_s \equiv \sqrt{T_e/m_i}$, the two conservation equations in toroidal coordinate system $(r, \vartheta, \varsigma)$ can be readily obtained for axisymmetric zonal flow on basis of Eqs.(B.13-B.15) by invoking $\partial/\partial \varsigma \to 0$, and $|\partial/\partial r| \gg (1/r)|\partial/\partial \vartheta|$, i.e. radial variation is much faster than poloidal variation. The leading order term of perpendicular viscosity is

$$\bar{\Pi}_{r\vartheta} = -m_i N \mu_B \frac{\partial \bar{u}_{\vartheta}}{\partial r}, \quad \text{(B.16)}$$

where $\bar{u}_{\vartheta} = \boldsymbol{e}_{\vartheta} \cdot \bar{\boldsymbol{u}}_{i,\perp}^{(0)}$, $\mu_B \equiv 3\nu_{ii}\rho_i^2/10$ is the classical perpendicular viscosity coefficient.

Charge conservation equation is

$$\left( \frac{\partial}{\partial \vartheta} - \frac{r}{R} \sin \vartheta \right) \frac{\sigma_{\delta}}{q^2} \frac{\partial}{\partial \vartheta} (\bar{n} - \bar{\varphi}) - 2\rho_s (1 + \tau_i) \sin \vartheta \frac{\partial \bar{n}}{\partial r}$$
$$- \frac{R\rho_s^2}{c_s} \frac{\partial}{\partial t} \frac{\partial^2}{\partial r^2} \bar{\varphi} + \mu_B \frac{R\rho_s^2}{c_s} \frac{\partial^4}{\partial r^4} (\bar{\varphi} + \tau_i \bar{n}) = \frac{R\rho_s}{c_s^2} \nabla \cdot \bar{\boldsymbol{u}}_{NP} \quad \text{(B.17)}$$

Particle conservation equation is

$$\frac{R^2}{c_s^2}\frac{\partial^2 \bar{\tilde{n}}}{\partial t^2} - \frac{(1+\tau_i)}{q^2}\left(\frac{\partial}{\partial \vartheta} - \frac{r}{R}\sin\vartheta\right)\frac{\partial \bar{\tilde{n}}}{\partial \vartheta} - \frac{2R\rho_s}{c_s}\sin\vartheta\frac{\partial}{\partial t}\frac{\partial}{\partial r}(\bar{\tilde{\varphi}} + \tau_i\bar{\tilde{n}})$$
$$-\frac{R^2\rho_s^2}{c_s^2}\frac{\partial^2}{\partial t^2}\nabla_\perp^2\bar{\tilde{\varphi}} + \mu_B\frac{R^2\rho_s^2}{c_s^2}\frac{\partial}{\partial t}\frac{\partial^4}{\partial r^4}(\bar{\tilde{\varphi}} + \tau_i\bar{\tilde{n}}) = \frac{R^2\rho_s}{c_s^3}\frac{\partial}{\partial t}\nabla\cdot\bar{\tilde{u}}_{NP} \quad . \tag{B.18}$$

In Eq.(B.17) $\sigma_\delta \equiv 2m_i c_s / m_e \nu_{ei} R$, $R$ is the major radius, $\sin\vartheta$ in the second term of Eq.(B.17) results from the geodesic curvature. In terms of $e\tilde{\phi}/T_e \to \tilde{\phi}$, the nonlinear polarization drift becomes $\bar{\tilde{u}}_{NP} = \left\langle \rho_s^2 c_s^2 (\boldsymbol{b}\times\nabla\tilde{\phi}\cdot\nabla)\nabla_\perp\tilde{\phi}\right\rangle_{en}$, where $\tilde{\phi}$ is 2D mode structure of micro-turbulence, which provides meso-scale driving force through three wave couplings, as depicted in Appendix C with the example of fluid ITG.

**Appendix C Poloidal moments of Reynolds stress**

In this appendix use is made of the methodology of [31] to calculate analytic formulae of poloidal moments of Reynolds stress induced by ITG micro-turbulence $\tilde{\phi}$

$$\Re[K] \equiv -\rho_s^2 c_s^2 \oint d\vartheta \left(\frac{\partial \tilde{\phi}}{r\partial \vartheta}\frac{\partial \tilde{\phi}}{\partial r}\right) K(\vartheta), \tag{C.1}$$

the same as Eq.(17) in section 2. For reader's convenience, the symbols are the same as those in [31] throughout this appendix, where the 2D mode structure is

$$\tilde{\phi}(r,\vartheta) = \sum_l |\tilde{\phi}_l(r)|\cos[S(r)-(m+l)\vartheta]. \tag{C.2}$$

The methodology used in [31] has been numerically verified by comparing the results of the analytic formula with the direct numerical integral of Eq.(C.1) for $K(\vartheta) := 1$ satisfactorily [31].

Substituting Eq.(C.2) into Eq.(C.1) yields

$$\Re[K] = -\rho_s^2 c_s^2 \sum_{l,l'} \oint d\vartheta \frac{(m+l)}{r}|\tilde{\phi}_l(r)|\sin[S(r)-(m+l)\vartheta]K(\vartheta)\times$$
$$\left\{\frac{d|\tilde{\phi}_{l'}(r)|}{dr}\cos[S(r)-(m+l')\vartheta] - |\tilde{\phi}_{l'}(r)|\frac{dS(r)}{dr}\sin[S(r)-(m+l')\vartheta]\right\} \quad . \tag{C.3}$$

This form will be simplified by defining two new functionals

$$P_{l,l'}[K] \equiv \oint d\vartheta \sin[S(r)-(m+l)\vartheta]\cos[S(r)-(m+l')\vartheta]K(\vartheta), \qquad (C.4)$$

$$Q_{l,l'}[K] \equiv \oint d\vartheta \sin[S(r)-(m+l)\vartheta]\sin[S(r)-(m+l')\vartheta]K(\vartheta). \qquad (C.5)$$

For micro-turbulence $m \gg 1$, integral of high order harmonic vanishes, Eq.(C.4) and (C.5) become

$$P_{l,l'}[K] = \frac{1}{2}\oint d\vartheta \sin[(l'-l)\vartheta]K(\vartheta), Q_{l,l'}[K] = \frac{1}{2}\oint d\vartheta \cos[(l-l')\vartheta]K(\vartheta). (C.6)$$

For k'th order sinusoidal harmonic

$$P_{l,l'}[\sin k\vartheta] = \frac{1}{4}(\delta_{l,l'-k} - \delta_{l,l'+k}), \quad Q_{l,l'}[\sin k\vartheta] = 0. \qquad (C.7)$$

For k'th order cosinoidal harmonic

$$P_{l,l'}[\cos k\vartheta] = 0, \quad Q_{l,l'}[\cos k\vartheta] = \frac{1}{4}(\delta_{l,l'-k} + \delta_{l,l'+k}). \qquad (C.8)$$

Substituting Eq.(C.4) and (C.5) into Eq.(C.3) yields

$$\Re[K] = -\rho_s^2 c_s^2 \sum_{l,l'} \frac{(m+l)}{r}\left\{|\tilde{\phi}_l(r)|\frac{\partial|\tilde{\phi}_{l'}(r)|}{\partial r}P_{l,l'}[K] - |\tilde{\phi}_l(r)||\tilde{\phi}_{l'}(r)|\frac{\partial S(r)}{\partial r}Q_{l,l'}[K]\right\}. (C.9)$$

Since $l/m \approx 1/\sqrt{n} \ll 1$ [31], for k'th cosinoidal and sinusoidal component, Eq.(C.9) can be written respectively as,

Cosinoidal moment of Reynolds stress:

$$\Re[\cos k\vartheta] = \frac{1}{4}\rho_s^2 c_s^2 k_\vartheta^2 \hat{s} \sum_{p=\pm 1}\sum_l \frac{dS(r)}{dx}|\tilde{\phi}_l(x)||\tilde{\phi}_{l+pk}(x)|. \qquad (C.10)$$

Sinusoidal moment of Reynolds stress:

$$\Re[\sin k\vartheta] = \frac{1}{4}\rho_s^2 c_s^2 k_\vartheta^2 \hat{s} \sum_{p=\pm 1} p \sum_l \frac{d|\tilde{\phi}_l(x)|}{dx}|\tilde{\phi}_{l+pk}(x)|. \qquad (C.11)$$

In Eq.(C.11) $x = k_\vartheta \hat{s}(r-r_j)$ is the dimensionless micro-radius in the ballooning theory [31].

Some factors in Eqs.(C.10), (C.11) can be found in [31]. They are

$$S(r) = \text{Re}\left[k_*(\lambda_*)(x-l) - \frac{1}{2\eta^2}(x-l)^2\right], \tag{C.12}$$

$$\left|\tilde{\phi}_l(r)\right| = \exp\left\{-\text{Im}\left[k_*(\lambda_*)(x-l) - \frac{1}{2\eta^2}(x-l)^2\right]\right\}\exp\left\{\text{Re}\left[-\frac{\text{n}}{2\beta_2}\left(\beta_1 + \frac{l}{\text{n}}\right)^2\right]\right\}, \tag{C.13}$$

$$\frac{dS(r)}{dx} = \frac{\text{Im}\left[\eta^2 k_*(\lambda_*)\right]}{\text{Im}\eta^2} + \frac{\text{Re}\eta^2}{2\text{Im}\eta^2}\frac{d\ln\left|\tilde{\phi}_l(x)\right|^2}{dx}, \tag{C.14}$$

$$I_m(x) \equiv \oint d\vartheta \left|\tilde{\phi}(r,\vartheta)\right|^2 = \frac{1}{2}\sum_l \left|\tilde{\phi}_l(x)\right|^2 \sim \Gamma(x). \tag{C.15}$$

In Eq.(C.15) '~' denotes that a constant factor before $\Gamma(x)$ is not important and neglected,

$$\Gamma(x) \equiv \sum_l \exp\left[-\bar{\alpha}(x-l-k_0)^2\right]\exp\left[-\bar{\beta}(l-x_\beta)^2\right]. \tag{C.16}$$

In Eq.(C.16) $\bar{\alpha} \equiv \text{Im}\eta^2/|\eta^2|^2$, $\bar{\beta} \equiv \text{Re}\beta_2/(\text{n}|\beta_2|^2)$, $k_0 \equiv -|\eta^2|^2 \text{Im}\bar{k}_*/\text{Im}\eta^2$, $\bar{k}_* \equiv k_*(\bar{\lambda}_*)$, $\bar{\lambda}_* \equiv -i\beta_1/\beta_2$, $\text{n} = |n|$, $x_\beta \equiv -\text{n}(\text{Re}\beta_1 + \text{Im}\beta_1 \text{Im}\beta_2/\text{Re}\beta_2)$.

Now, we are ready to calculate Eqs.(C.10), (C.11). Use is made of Eq.(C.13) to obtain

$$\frac{\left|\tilde{\phi}_{l+pk}(x)\right|}{\left|\tilde{\phi}_l(x)\right|} = \exp\left\{pk\left[\bar{\alpha}(x-l) + \bar{\beta}(x_\beta - l) + \text{Im}k_*(\lambda_*)\right] - \frac{(\bar{\alpha}+\bar{\beta})k^2}{2}\right\}. \tag{C.17}$$

From Eq.(C.15) and (C.16), we obtain

$$\left|\tilde{\phi}_l(x)\right|^2 \sim 2\exp\left[-\bar{\alpha}(x-l-k_0)^2\right]\exp\left[-\bar{\beta}(l-x_\beta)^2\right]. \tag{C.18}$$

Multiplying Eq.(C.17) by (C.18) yields

$$\sum_l \left|\tilde{\phi}_l(x)\right|\left|\tilde{\phi}_{l+pk}(x)\right| \sim 2\Gamma_{p,k}(x)\exp\left\{pk\left[\bar{\alpha}k_0 + \text{Im}k_*(\lambda_*)\right] - \frac{1}{4}(\bar{\alpha}+\bar{\beta})k^2\right\}, \tag{C.19}$$

with

$$\Gamma_{p,k}(x) \equiv \sum_l \exp\left[-\bar{\alpha}(x-l-k_p)^2\right]\exp\left[-\bar{\beta}(x_{\beta,p}-l)^2\right]. \tag{C.20}$$

In Eq.(C.20) $k_p \equiv k_0 + pk/2$, $x_{\beta,p} \equiv x_\beta - pk/2$. Eq.(C.20) can also be written as

$$\Gamma_{p,k}(x) = \exp\left[-\frac{\bar{\alpha}\bar{\beta}}{(\bar{\alpha}+\bar{\beta})}(x - x_{\beta,p} - k_p)^2\right]\Pi_{p,k}(x). \tag{C.21}$$

In Eq.(C.21)

$$\Pi_{p,k}(x) \equiv \sum_{l}\exp\left\{-(\bar{\alpha}+\bar{\beta})\left[\frac{\bar{\alpha}}{\bar{\alpha}+\bar{\beta}}x - l - \frac{\bar{\alpha}k_p - \bar{\beta}x_{\beta,p}}{(\bar{\alpha}+\bar{\beta})}\right]^2\right\}. \tag{C.22}$$

The function $\Pi_{p=0,k}(x)$ has been discussed in the Appendix B in [31] in detail, and is shown close to a constant for $(\bar{\alpha}+\bar{\beta}) < 2$. This is still true for other integer $p$, because the translational invariance of this function is independent of $p$. Neglecting the small variation of $\Pi_{p,k}(x)$, Eq.(C.19) becomes

$$\sum_{l}|\tilde{\phi}_l(x)||\tilde{\phi}_{l+pk}(x)| \approx 2I_m(x_0)\exp\left[-\frac{(x-x_0)^2}{n\sigma}\right] \times \\ \exp\left\{pk[\bar{\alpha}k_0 + \mathrm{Im}\, k_*(\lambda_*)] - \frac{1}{4}(\bar{\alpha}+\bar{\beta})k^2\right\} \tag{C.23}$$

In Eq.(C.23) $x_0 \equiv x_\beta + k_0$, $\sigma \equiv (\bar{\alpha}+\bar{\beta})/(n\bar{\alpha}\bar{\beta}) = |\beta_2|^2/\mathrm{Re}\,\beta_2 + |\eta|^2/(n\,\mathrm{Im}\,\eta^2)$.

Substituting Eq.(C.18) into (C.14) yields

$$\frac{dS(x)}{dx} = \frac{\mathrm{Im}[\eta^2 k_*(\lambda_*)]}{\mathrm{Im}\,\eta^2} - \frac{\mathrm{Re}\,\eta^2}{\mathrm{Im}\,\eta^2}\bar{\alpha}(x - l - k_0), \tag{C.24}$$

in which $\bar{\alpha}(x - l - k_0)$ can be equivalently replaced by the operator $(-d/dx + pk\bar{\alpha})/2$,

$$\sum_{l}\frac{dS(x)}{dx}|\tilde{\phi}_l(x)||\tilde{\phi}_{l+pk}(x)| = \\ \left\{\frac{\mathrm{Im}[\eta^2\langle k_*(\lambda_*)\rangle]}{\mathrm{Im}\,\eta^2} + \frac{\mathrm{Re}\,\eta^2}{2\,\mathrm{Im}\,\eta^2}\left(\frac{d}{dx} - pk\bar{\alpha}\right)\right\}\sum_{l}|\tilde{\phi}_l(x)||\tilde{\phi}_{l+pk}(x)| \tag{C.25}$$

In Eq.(C.25) $\langle k_*(\lambda_*)\rangle \equiv \sum_l k_*(\lambda_*)|\tilde{\phi}_l(x)||\tilde{\phi}_{l+pk}(x)| / \sum_l |\tilde{\phi}_l(x)||\tilde{\phi}_{l+pk}(x)|$. As discussed in [31], $\langle k_*(\lambda_*)\rangle$ is a slowly varying function of radius, it can be further approximated by a simple constant $k_*(x_*)$ with $x_* \equiv -i(\beta_1 + x_0/n)/\beta_2$.

Substituting Eq.(B.23) into (B.25), then into Eq.(C.10), the analytic expression of cosinoidal moment of Reynolds stress is found to be

$$\Re[\cos k\vartheta] = \frac{1}{2}\rho_s^2 c_s^2 k_g^2 \hat{s} I_m(x_0) \sum_{p=\pm 1} \left\{ \frac{\operatorname{Im}[\eta^2 k_*(x_*)]}{\operatorname{Im}\eta^2} - \frac{\operatorname{Re}\eta^2}{\operatorname{Im}\eta^2} \frac{(x-x_0)}{n\sigma} - \frac{\operatorname{Re}\eta^2}{2|\eta^2|^2} pk \right\} \times$$

$$\exp\left[-\frac{(x-x_0)^2}{n\sigma}\right] \exp\left\{ pk[\bar{\alpha}k_0 + \operatorname{Im} k_*(x_*)] - \frac{1}{4}(\bar{\alpha}+\bar{\beta})k^2 \right\}. \quad \text{(C.26)}$$

For $k = 0$, Eq.(B.26) becomes

$$\Re[1] = \rho_s^2 c_s^2 k_g^2 \hat{s} I_m(x_0) \left\{ \frac{\operatorname{Im}[\eta^2 k_*(x_*)]}{\operatorname{Im}\eta^2} - \frac{\operatorname{Re}\eta^2}{\operatorname{Im}\eta^2} \frac{(x-x_0)}{n\sigma} \right\} \exp\left[-\frac{(x-x_0)^2}{n\sigma}\right], \quad \text{(C.27)}$$

which is the same result as Eq.(42) in [31].

The analytic expression of sinusoidal moment of Reynolds stress can be obtained similarly as

$$\Re[\sin k\vartheta] = \frac{1}{2}\rho_s^2 c_s^2 k_g^2 \hat{s} I_m(x_0) \sum_{p=\pm 1} p\left[-\frac{(x-x_0)}{n\sigma} - \frac{p\bar{\alpha}k}{2}\right] \exp\left[-\frac{(x-x_0)^2}{n\sigma}\right] \times$$

$$\exp\left\{ pk[\bar{\alpha}k_0 + \operatorname{Im} k_*(x_*)] - \frac{1}{4}(\bar{\alpha}+\bar{\beta})k^2 \right\}. \quad \text{(C.28)}$$

Both the cosinoidal and sinusoidal moment are composed of a monopole and a dipole. However, the dipole is much larger than the monopole for cosinoidal moment, since $k_*(x_*)$ is small. But for sinusoidal moment, the monopole structure looks more apparent.

## Appendix D TLFZF equation and GAM dispersion

As stated in the introductory section, the framework of the present paper in studying GAM is the extension of equations of [11] to include toroidal coupling of the first harmonic sinusoidal component of sound wave owing to geodesic curvature. A question may arise from the concerns how far the toroidal coupling would modify the model used in [11]. Whether or not so many feathers derived in [11] regarding LFZF might survive etc. The quick answer is that in the low frequency limit the toroidal effect on the TLFZF branch is simply a quantitative change of the inertia of zonal flow as shown below. Since a free parameter $a_{neo}$ has been introduced in [11] to get the theory suitable for various models of zonal flow inertia, the results obtained in [11] are still valid.

Eqs.(18), (19) can be merged into a single equation by eliminating $\chi_1^{(s)}$

$$\begin{aligned}
&\left[\frac{\partial^2}{\partial t^2}\left(1+D(\tau_e)\rho_i^2\frac{\partial^2}{\partial r^2}\right)+2(1+\tau_i)\frac{\psi c_s^2}{R^2}\left(1+\frac{1}{2q^2}\right)+\mu(1+\tau_i)\rho_s^2\frac{\partial}{\partial t}\frac{\partial^4}{\partial r^4}\right]\times \\
&\quad\left[\frac{\partial \bar{\upsilon}}{\partial t}-\mu\frac{\partial^2 \bar{\upsilon}}{\partial r^2}+\frac{1}{2}\frac{\partial}{\partial r}\left\{\left(\tilde{\mathfrak{R}}[1]+\varepsilon\tilde{\mathfrak{R}}[\cos\vartheta]\right)\cos 2\Theta\right\}\right] = \\
&-2\frac{\psi c_s^2}{R^2}(1+\tau_i)\left[\mu\frac{\partial^2 \bar{\upsilon}}{\partial r^2}-\frac{1}{2}\frac{\partial}{\partial r}\left\{\left(\tilde{\mathfrak{R}}[1]+\varepsilon\tilde{\mathfrak{R}}[\cos\vartheta]\right)\cos 2\Theta\right\}\right] \\
&-\frac{\psi\rho_s c_s}{R}(1+\tau_i)\frac{\partial}{\partial t}\frac{\partial^2}{\partial r^2}\left\{\left(\tilde{\mathfrak{R}}[\sin\vartheta]+\frac{\varepsilon}{3}\tilde{\mathfrak{R}}[\sin 2\vartheta]\right)\cos 2\Theta\right\}
\end{aligned} \qquad \text{(D.1)}$$

In low frequency limit, the temporal derivative in the first term can be neglected, which reduces Eq.(D.1) to be

$$\begin{aligned}
&(1+2q^2)\frac{\partial \bar{\upsilon}}{\partial t}-\mu\frac{\partial^2 \bar{\upsilon}}{\partial r^2}+\frac{1}{2}\frac{\partial}{\partial r}\left\{\left(\tilde{\mathfrak{R}}[1]+\varepsilon\tilde{\mathfrak{R}}[\cos\vartheta]\right)\cos 2\Theta\right\}= \\
&-q^2R\frac{\rho_s}{c_s}\frac{\partial}{\partial t}\frac{\partial^2}{\partial r^2}\left\{\left(\tilde{\mathfrak{R}}[\sin\vartheta]+\frac{\varepsilon}{3}\tilde{\mathfrak{R}}[\sin 2\vartheta]\right)\cos 2\Theta\right\}
\end{aligned} \qquad \text{(D.2)}$$

Since the r.h.s. of Eq.(D.2) is associated with a low frequency, and the toroidal correction is associated with a small parameter $\varepsilon$, we readily obtain the final equation for the low frequency branch by neglecting these small quantitative corrections

$$(1+2q^2)\frac{\partial \bar{\upsilon}}{\partial t}-\mu\frac{\partial^2 \bar{\upsilon}}{\partial r^2}+\frac{1}{2}\frac{\partial}{\partial r}\left\{\tilde{\mathfrak{R}}[1]\cos 2\Theta\right\}=0. \qquad \text{(D.3)}$$

It is precisely the leading order equation of TLFZF under fluid model in simple tokamak configuration. Compared to Eq.(19) of [11], $a_{neo}:=1+2q^2$.

The GAM dispersion relation can be obtained from left hand side of Eq.(D.1)

$$\left\{\frac{\partial^2}{\partial t^2}\left[1+D(\tau_e)\rho_i^2\frac{\partial^2}{\partial r^2}\right]+2(1+\tau_i)\frac{\psi c_s^2}{R^2}\left(1+\frac{1}{2q^2}\right)+\mu(1+\tau_i)\rho_s^2\frac{\partial}{\partial t}\frac{\partial^4}{\partial r^4}\right\}\bar{\upsilon}=0 \,. \quad \text{(D.4)}$$

In the local Fourier representation

$$\omega^2=\omega_G^2\left[1+D(\tau_e)\rho_i^2 k^2\right],\quad \omega_G^2=2(1+\tau_i)\frac{\psi c_s^2}{R^2}\left(1+\frac{1}{2q^2}\right), \quad \text{(D.5)}$$

where use is made of the condition that $D(\tau_e)\rho_i^2 k^2 \ll 1$ and terms $\sim O(k^2)$ are retained.

## Acknowledgments

The authors would like to acknowledge Dr. D. F. Kong, X. H. Zhang and M. Y. Wang for helpful discussion on experimental observations. The present work was supported in part by the National MCF Energy R&D Program under Grant Nos. 2018YFE0311200 and 2017YFE0301204, the Natural Science Foundation of China under Grant Nos. No. U1967206 and 11975231, National Natural Science Foundation of China (NSFC-11805203, 11775222), Key Research Program of Frontier Science CAS (QYZDB-SSW-SYS004) and the U.S. Dept. of Energy Grant No. DE -FG02-04ER-54742.